\documentclass[AMA,STIX1COL]{WileyNJD-v2}

\usepackage{dirtree}
\usepackage{fancyvrb}
\usepackage{xcolor} 
\usepackage{xspace}

\newcommand{\toskerContainer}{\textit{tos\-ker.no\-des.Con\-tai\-ner}\xspace}
\newcommand{\toskerSoftware}{\textit{tos\-ker.no\-des.Soft\-wa\-re}\xspace}
\newcommand{\toskerVolume}{\textit{tos\-ker.no\-des.Vo\-lu\-me}\xspace}

\newcommand{\toscaHostedOn}{\textit{tos\-ca.re\-la\-tion\-ships.Hos\-ted\-On}\xspace}
\newcommand{\toscaConnectsTo}{\textit{tos\-ca.re\-la\-tion\-ships.Con\-nects\-To}\xspace}
\newcommand{\toscaAttachesTo}{\textit{tos\-ca.re\-la\-tion\-ships.At\-ta\-ches\-To}\xspace}
\newcommand{\toscaDependsOn}{\textit{tos\-ca.re\-la\-tion\-ships.De\-pends\-On}\xspace}

\newcommand{\toskose}{\textsc{Tos\-kose}\xspace}
\newcommand{\toskoseunit}{\toskose \textsc{Unit}\xspace}
\newcommand{\toskosemanager}{\toskose \textsc{Manager}\xspace}
\newcommand{\toskosepackager}{\toskose \textsc{Packager}\xspace}

\newcommand{\manager}{\textit{Manager}\xspace}
\newcommand{\packager}{\textit{Packager}\xspace}
\newcommand{\unit}{\textit{Unit}\xspace}
\newcommand{\sone}{\textit{s1}\xspace}
\newcommand{\stwo}{\textit{s2}\xspace}
\newcommand{\sthree}{\textit{s3}\xspace}
\newcommand{\cone}{\textit{c1}\xspace}
\newcommand{\ctwo}{\textit{c2}\xspace}
\newcommand{\cthree}{\textit{c3}\xspace}
\newcommand{\cfour}{\textit{c4}\xspace}

\newcommand{\thinking}{\textit{Thinking}\xspace}
\newcommand{\gui}{\textit{GUI}\xspace}
\newcommand{\node}{\textit{Node}\xspace}
\newcommand{\logsniffer}{\textit{Log\-sniffer}\xspace}
\newcommand{\api}{\textit{API}\xspace}
\newcommand{\apiSoftware}{\textit{tos\-ker.no\-des.API\-Soft\-wa\-re}\xspace}
\newcommand{\pushdefault}{\textit{push\_de\-fault}\xspace}
\newcommand{\maven}{\textit{Maven}\xspace}
\newcommand{\mongodb}{\textit{MongoDB}\xspace}
\newcommand{\dbvolume}{\textit{MongoDB}\xspace}

\newcommand{\eg}{e.g.,\xspace}
\newcommand{\ie}{i.e.,\xspace}
\newcommand{\figscale}{.35}

\newcommand{\clwidth}{.95\textwidth} 

\articletype{Research Article}%

\begin{document}

\title{Component-aware Orchestration of Cloud-based Enterprise Applications, from TOSCA to Docker and Kubernetes}

\author[1]{Matteo Bogo}

\author[1]{Jacopo Soldani*}

\author[1]{Davide Neri}

\author[1]{Antonio Brogi}

\authormark{Bogo M \textsc{et al}}

\address{\orgdiv{Department of Computer Science}, \orgname{University of Pisa}, \orgaddress{\state{Pisa}, \country{Italy}}}



\corres{*Jacopo Soldani, c/o Dipartimento di Informatica, Universit\`a di Pisa, Largo B. Pontecorvo 3, 56127, Pisa, Italia. \email{soldani@di.unipi.it}}

\abstract[Summary]{%
	Enterprise IT is currently facing the challenge of coordinating the management of complex, multi-component applications across heterogeneous cloud platforms. 
	Containers and container orchestrators provide a valuable solution to deploy multi-component applications over cloud platforms, by coupling the lifecycle of each application component to that of its hosting container.
	We hereby propose a solution for going beyond such a coupling, based on the OASIS standard TOSCA and on Docker.
	We indeed propose a novel approach for deploying multi-component applications on top of existing container orchestrators, which allows to manage each component independently from the container used to run it. We also present prototype tools implementing our approach, and we show how we effectively exploited them to carry out a concrete case study.
}

\keywords{application orchestration, cloud, Docker, TOSCA}

\jnlcitation{\cname{%
 \author{M. Bogo},
 \author{J. Soldani},
 \author{D. Neri}, and
 \author{A. Brogi}}, (\cyear{2019}), 
 \ctitle{Component-aware Orchestration of Cloud-based Enterprise Applications, from TOSCA to Docker and Kubernetes}, \cjournal{Submitted for publication}. 
}

\maketitle


\section{Introduction}
\label{sec:intro}
Cloud computing is a flexible, cost-effective and proven delivery platform for running on-demand distributed applications\cite{cloud-computing-concepts}.
To fully exploit the potentials of cloud computing and facilitate the scalability, reliability and portability of applications, various cloud-native architectural styles have emerged (with microservices being one of the most prominent examples).
This has resulted in a growth of the complexity of applications, which nowadays integrate dozens (if not hundreds) of interacting components\cite{pains-gains-microservices}.
The problem of automating the deployment and management of such complex, multi-component applications over heterogeneous cloud platforms has hence become one of the main challenges in enterprise IT\cite{pattern-based-multi-cloud-migration,cloud-systems-challenges}.

The components forming an application are typically deployed on cloud platforms by relying on virtualisation technologies. 
Container-based virtualisation\cite{container-based-virtualisation} is getting more and more momentum in this scenario, as it provides an isolated and lightweight virtual runtime environment\cite{cloud-container-technologies}.
Docker (\url{https://www.docker.com}) constitutes the \textit{de-facto} standard for container-based virtualisation, and it permits packaging software components (together with all software dependencies they need to run) in Docker images, which are then
exploited as read-only templates to create and run Docker containers.
Container orchestrators are then used to automate the deployment and management of containerised applications at a large scale.
Docker Swarm (\url{https://docs.docker.com/engine/swarm}) and Kubernetes (\url{https://kubernetes.io}) are currently the most popular solutions for orchestrating Docker  containers, providing  all necessary abstractions for scaling, discovering, load-balancing and interconnecting Docker containers over single and multi-host systems\cite{docker-up-running,kubernetes-up-running}.

Docker containers are however treated as a sort of "black-boxes", since they constitute the minimal entity that can be orchestrated.
Container orchestrators can indeed create, start, stop and delete containers, but they do not provide support for coordinating the management of the components running within containers.
The lifecycle of the software components forming a containerised application is hence tightly coupled to that of their hosting containers.
For instance, when the orchestrator creates and starts a container, all the software components it contains have to be created and started as well, as the orchestrator does not provide a support for creating or starting them afterwards.
The same holds when containers are stopped or deleted, as the components they are hosting get stopped and deleted as well.
This is because currently existing container orchestrators do not provide a support for coordinating the management of software components independently from that of their hosting containers\cite{tosker}.

Decoupling the management of application components from that of the containers hosting them can anyway bring various advantages.
For instance, this allows to employ Docker containers as so-called \textit{system containers}, \ie a lightweight portable alternative to virtual machines\cite{enterprise-docker}. 
It also enables inter-process communication within the components running in the same container, without requiring them to necessarily communicate over the network\cite{ipc-containers}.
This helps saving resources, thus containing costs or enabling a fine-grained orchestration on infrastructures where computing and networking resources are costly and limited, \eg in edge clusters\cite{containers-edge}. 
Other known advantages of decoupling application components from their hosting containers are maintainability and reuse\cite{toskeriser}.
Deployment requirements of multi-component applications change over time.
If components are coupled to their hosting containers, this requires to re-package them from scratch whenever their deployment requirements change.

{Following this idea, we hereafter} propose a solution for managing the software components forming an application independently from the containers used to run them. 
The proposed solution is intended to allow a more flexible, component-aware management of multi-component applications on top of existing Docker-based container orchestrators, and it does so by relying on the OASIS standard TOSCA for specifying and orchestrating multi-component applications.
More precisely, starting from an existing representation of multi-component applications in TOSCA (taken from our previous work\cite{tosker}), we provide the following contributions:
\begin{itemize}
    \item We propose a novel approach for managing the lifecycle of software components forming a multi-component application independently from that of the containers used to host such components.
    \item We present {the prototype} tools implementing our approach. These include a service enabling the component-aware runtime management of multi-component applications and a packager for generating the deployment artifacts needed to ship and manage applications on existing Docker-based container orchestrators.
    \item We showcase the effectiveness of our approach and prototype tools by reporting on how we exploited them to run a concrete case study based on an existing benchmark application.
\end{itemize}

It is worth highlighting that our solution is not intended to implement a new orchestrator from scratch, as for instance we did with the TosKer orchestration engine\cite{tosker}.
We instead aim at enabling a component-aware management of multi-component applications on top of existing, production-ready container orchestrator (e.g.,~Docker Swarm or Kubernetes), in order to exploit all features they already provide, e.g.,~multi-host deployment and network overlays.
In this way we do not need to re-implement such features from scratch, and we can hence focus on seamlessly extending the orchestration support they provide.

\smallskip \noindent
The rest of this paper is organised as follows. 
Sects.~\ref{sec:background} and \ref{sec:big-picture} provides some background and a bird-eye view of the proposed approach, respectively.
Sects.~\ref{sec:toskose-unit}, \ref{sec:toskose-manager} and \ref{sec:packaging} then introduce the open-source {prototype} tools implementing our approach.
Sect.~\ref{sec:case-studies} shows how we exploited our approach to run a concrete case study based on an existing multi-component application.
Finally, Sect.~\ref{sec:related} and \ref{sec:conclusions} discuss related work and draw some concluding remarks, respectively.

\section{Background}
\label{sec:background}
\subsection{TOSCA}
\label{ssec:tosca}

TOSCA\cite{tosca} (\textit{Topology and Orchestration Specification for Cloud Applications}) is an OASIS standard whose main goals are to enable (i) the specification of portable cloud applications and (ii) the automation of their deployment and management. 
TOSCA provides a YAML-based and machine-readable modelling language that allows to describe cloud applications.
Obtained specifications can then be processed to automate the deployment and management of the specified applications.
%

\begin{figure}
\centering
\includegraphics[scale=\figscale]{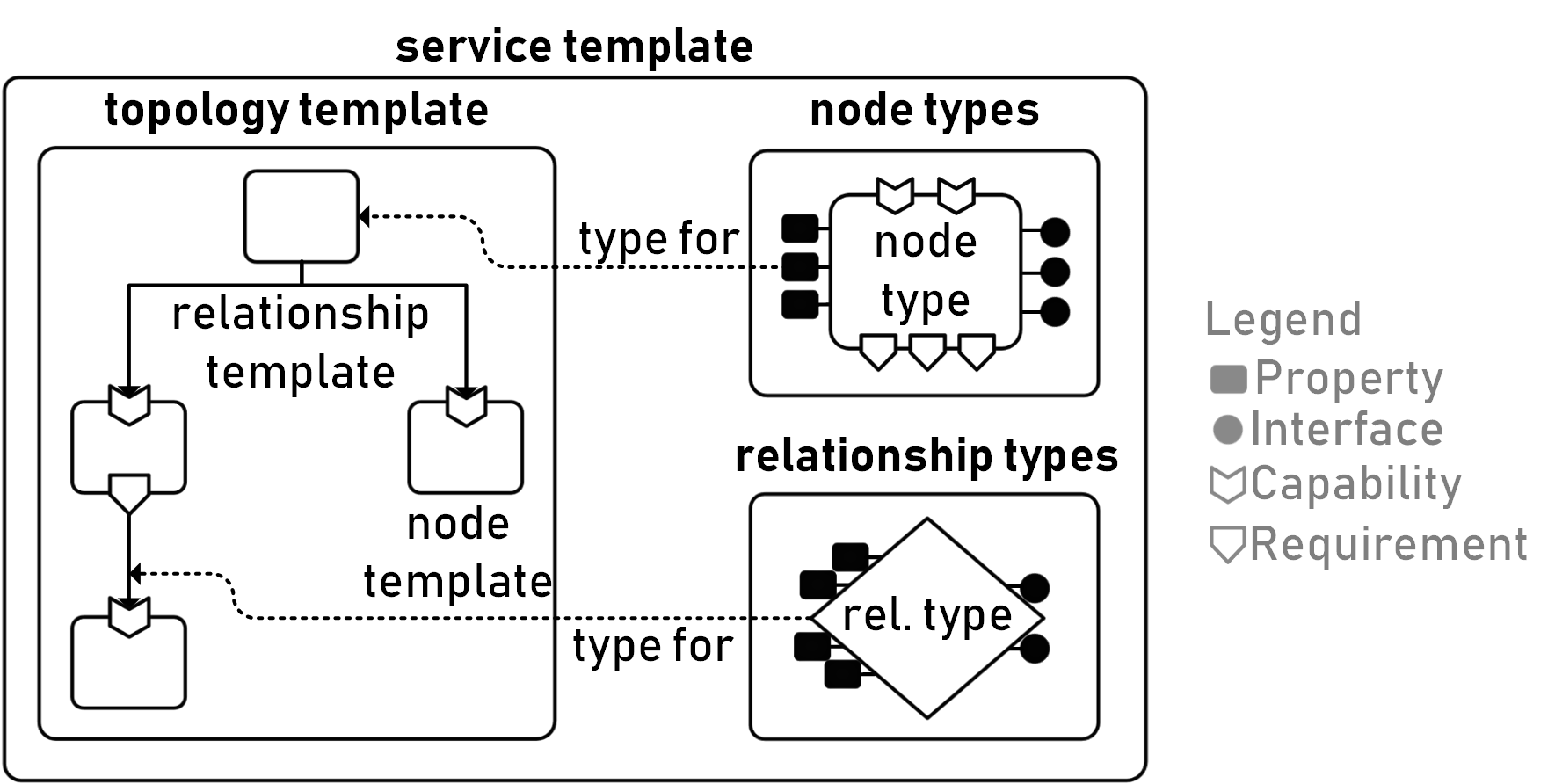}
\caption{The TOSCA metamodel~\cite{tosca}.}
\label{fig:service-template}
\end{figure}

TOSCA allows to specify a cloud application as a service template, that is in turn composed by a topology template, and by the types needed to build such a topology template (Fig.~\ref{fig:service-template}).
The topology template is a typed directed graph that describes the topological structure of a multi-component application.~Its nodes (called node templates) model the application components, while its edges (called relationship templates) model the relations occurring among such components. 
%

Node templates and relationship templates are typed by means of node types and relationship types, respectively.
A node type defines the observable properties of a component, its possible requirements, the capabilities it may offer to satisfy other components' requirements, and the interfaces through which it offers its management operations.
Requirements and capabilities are also typed, to permit specifying the properties characterising them.
A relationship type instead describes the observable properties of a relationship occurring between two application components. 
As the TOSCA type system supports inheritance, a node/relationship type can be defined by extending another, thus allowing the former to inherit the latter's properties, requirements, capabilities, interfaces, and operations (if any).

Node templates and relationship templates also specify the artifacts needed to actually realise their deployment or to implement their management operations. 
As TOSCA allows artifacts to represent contents of any type (e.g.,~scripts, executables, images, configuration files, etc.), the metadata needed to properly access and process them is described by means of artifact types.

Finally, to enable their actual deployment, TOSCA applications are packaged and distributed in CSARs (\textit{Cloud Service ARchives}).
A CSAR is a zip archive containing an application specification along with the concrete artifacts realising the deployment and management operations of its components.

\subsection{Modelling Multi-component, Docker-based Applications with TOSCA}
\label{ssec:bg-tosker-types}
\begin{figure}[t]
\centering
\begin{minipage}{.32\textwidth}
    \centering
    \includegraphics[scale=\figscale]{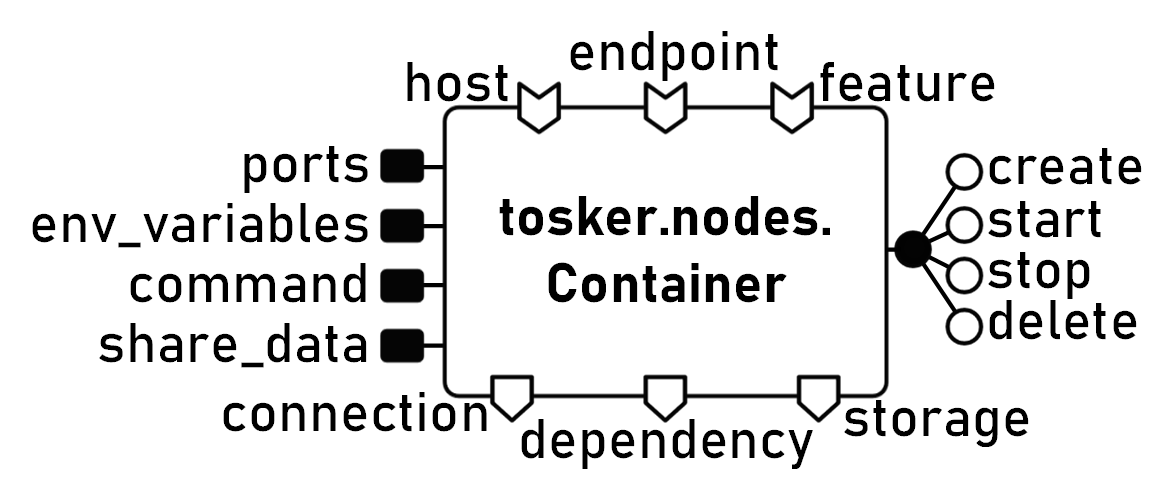}
\end{minipage}
\begin{minipage}{.32\textwidth}
    \centering
    \includegraphics[scale=\figscale]{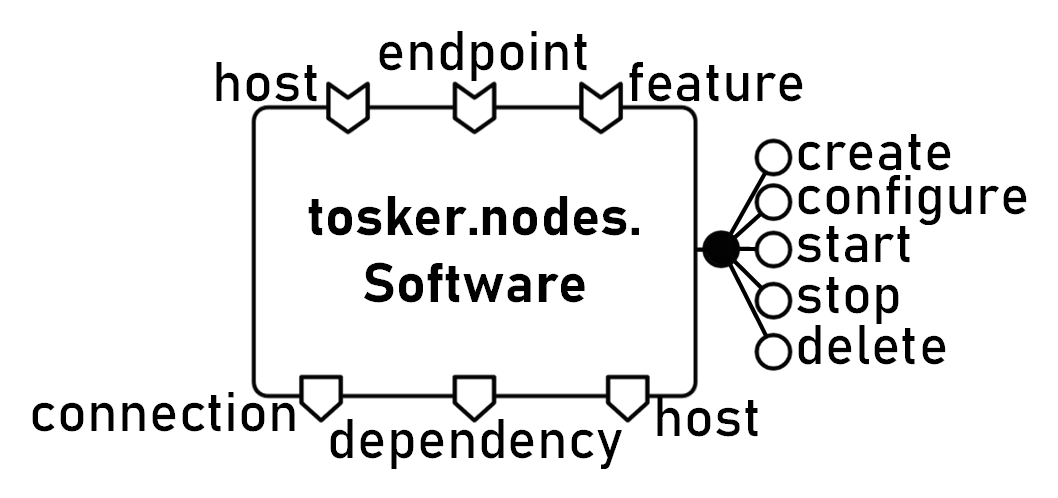}
\end{minipage}
\begin{minipage}{.32\textwidth}
    \centering
    \includegraphics[scale=\figscale]{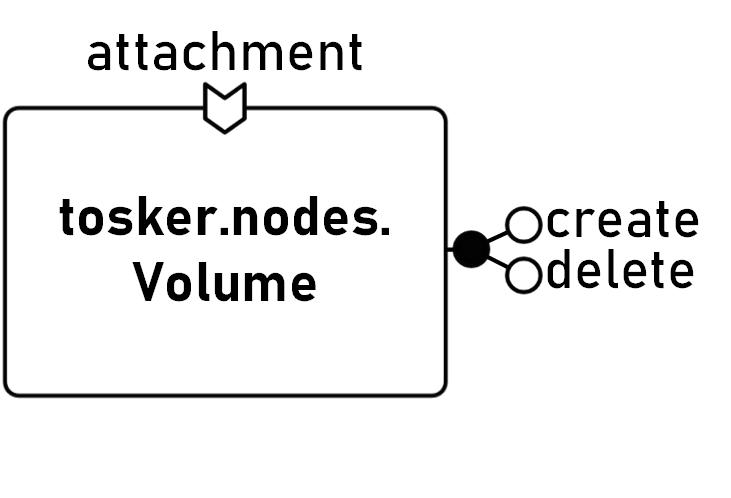}
\end{minipage}
\caption{TOSCA node types for multi-component, Docker-based applications, viz.,~\toskerContainer, \toskerSoftware, and \toskerVolume.}
\label{fig:tosker-types}
\end{figure}

In our previous work\cite{tosker}, we defined a TOSCA-based representation for specifying the software components forming an application, as well as the Docker containers and Docker volumes used to form their runtime infrastructure.
More precisely, three different TOSCA node types (Fig.~\ref{fig:tosker-types}) allow to distinguish among the Docker containers, Docker volumes and software components forming a multi-component application.
\begin{itemize}
\item \toskerContainer allows to describe Docker containers, by indicating whether a container requires a \textit{connection} (to another Docker container or  to an application component), whether it has a generic \textit{dependency} on another node in the topology, or whether it needs some persistent \textit{storage} (hence requiring to be attached to a Docker volume).
\toskerContainer also permits indicating whether a container can \textit{host} an application component, whether it offers an \textit{endpoint} where to connect to, or whether it offers a generic \textit{feature} (to satisfy a generic \textit{dependency} requirement of another container/application component).
To complete the description, \toskerContainer can contain properties (\textit{ports}, \textit{env\_va\-ri\-ab\-les}, \textit{command}, and \textit{share\_data}, respectively) for specifying the port mappings, the environment variables, the command to be executed when running the corresponding Docker container, the list of files and folders to share with the host.
Finally, \toskerContainer lists the operations to manage a container (which corresponds to the basic operations offered by the Docker platform to manage Docker containers~\cite{docker-up-running}).
\item \toskerVolume allows to specify Docker volumes, and it defines a capability \textit{attachment} to indicate that a Docker volume can satisfy the \textit{storage} requirements of Docker containers.
It also lists the operations to manage a Docker volume (which corresponds to the operations to create and delete Docker volumes offered by the Docker platform~\cite{docker-up-running}).
\item \toskerSoftware allows to represent the software components forming a multi-component application. 
It allows to specify whether an application component requires a \textit{connection} (to a Docker container or to another application component), whether it has a generic \textit{dependency} on another node in the topology, and that it has to be \textit{host}ed on a Docker container or on another component. 
\toskerSoftware also permits indicating whether an application component can \textit{host} another application component, whether it provides an \textit{endpoint} where to connect to, or whether it offers a generic \textit{feature} (to satisfy a generic \textit{dependency} requirement of a container/application component).
Finally, \toskerSoftware lists the operations to manage an application component by exploiting the TOSCA standard lifecycle interface~\cite{tosca} (viz.,~\textit{create}, \textit{configure}, \textit{start}, \textit{stop},  \textit{delete}).
\end{itemize}
The interconnections and interdependencies among the nodes forming a multi-component application can instead be specified by exploiting the TOSCA normative relationship types~\cite{tosca}.
The relationship type \toscaAttachesTo allows to attach a Docker volume to a Docker container.
\toscaConnectsTo allows to describe the network connections to establish between Docker containers and/or application components.
\toscaHostedOn allows to indicate that an application component is hosted on another component or on a Docker container (e.g.,~to indicate that a web service is hosted on a web server, which is in turn hosted on a Docker container).
Finally, \toscaDependsOn allows to represent generic dependencies between the nodes of a multi-component application (e.g.,~to denote that a component must be deployed before another, as the latter depends on the availability of the former to properly work).

Concrete examples of modelling of multi-component applications with the above recapped TOSCA representation can be found in Sect.~\ref{sec:case-studies}.

\section{Bird-eye view of our approach}
\label{sec:big-picture}
\label{sec:toskose}
Our ultimate objective is to enable a component-aware management of multi-component applications by piggybacking on existing container orchestrators (such as Docker Swarm and Kubernetes). 
We indeed aim at seamlessly extending the support they provide for orchestrating containers, so that the lifecycle of application components is not entangled to that of their hosting containers, but rather allowing components to be managed independently.

\begin{figure}
  \centering
  \includegraphics[scale=\figscale]{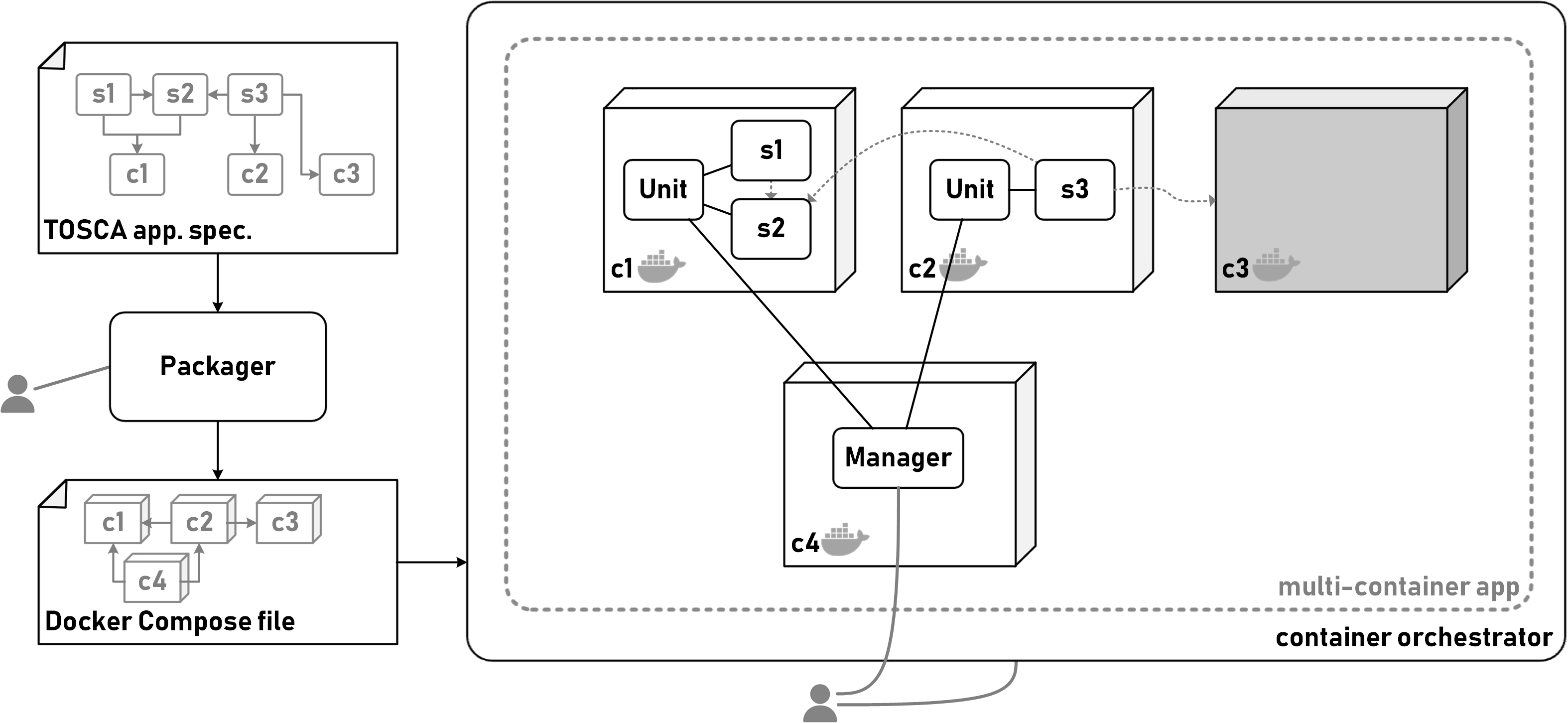}
  \caption{Bird-eye view of our approach for enabling a component-aware orchestration of the management multi-component applications on top of existing container orchestrators.}
  \label{fig:big-picture}
\end{figure}

Fig.~\ref{fig:big-picture} provides an high-level overview on the orchestration approach we propose.
The input is the TOSCA specification of a multi-component application.
In the figure, the considered application is composed by three services (\ie \sone, \stwo and \sthree) and three containers (\ie \cone, \ctwo and \cthree).
Containers \cone and \ctwo are used as system containers\cite{enterprise-docker}, \ie as lightweight virtual environments for hosting services \sone and \stwo and service \sthree, respectively.
The container \cthree is instead a standalone container running its default main process.
Services \sone and \sthree also connect to \stwo and to \stwo and \cthree to deliver their businesses.

Given such a TOSCA application specification, a \packager generates the Docker Compose file allowing to deploy the application on top of Docker-compliant container orchestrators.
The obtained artifact not only packages the components forming an application within the containers hosting them, but also seamlessly extends the application deployment by introducing some additional components enabling the desired component-aware orchestration.
\begin{itemize}
	\item A {\unit} is included in each container packaging some application component {(\ie \cone and \ctwo)}. 
	{Each \unit} is responsible of managing the lifecycle of the components packaged within the container it appears in, by launching the concrete artifacts (\eg bash scripts) implementing their management operations.
	\item A {\manager} is included and packaged within a {newly added container (\ie \cfour)}.
	The {\manager} is responsible of interacting and coordinating the {\unit}s, in order to manage the components forming an application, and based on the commands issued by an application administrator.
\end{itemize}

With our approach, an application administrator can continue to deploy and manage the containers in an application by exploiting existing container orchestrators.
She can then deploy and manage each software component forming her application by issuing commands to the containerised {\manager}, which will then properly forward it to the {\unit}s to enact the corresponding management operation.
For instance, once the containers in Fig.~\ref{fig:big-picture} are all running, an application administrator may issue the commands to install and start service \sone to the {\manager}.
The latter will forward the corresponding requests to the {\unit} in the container \cone, which will launch the bash scripts implementing the to-be-enacted management operations.

It is worth noting that our approach seamlessly integrates with any Docker-compliant container orchestrator, as the newly introduced components (\ie the {\manager} and the {\unit}s) are themselves packaged within Docker containers.
In addition, by piggybacking on existing container orchestrator, our approach allows to uniformly manage containers, independently on whether they are used as lightweight virtual hosting environments for application components (as \cone e \ctwo in the figure) or whether they are used as standalone services (as \cthree in the figure).
This allows application administrators to choose whether to enable the fine-grain management of components or to couple their lifecycle to the containers they appear in.
In the former case, she has to distinguish components and containers in the TOSCA application specification (and provide the artifacts implementing the lifecycle operations of such components), while in the latter case she just packages the components within the corresponding containers (by writing proper Dockerfiles --- as she would be doing with current Docker-based orchestrators).

\smallskip \noindent
In the following, we illustrate how we concretely obtained the above illustrated orchestration solution, by designing and developing the \toskose open-source {toolset}. 
We first show the issues and design choices for allowing \textit{{\unit}s} to independently manage the software components packaged within a container (Sect.~\ref{sec:toskose-unit}).
We then show how we designed and developed the {\manager} to enable the orchestration of multi-component applications on top of existing orchestrators (Sect.~\ref{sec:toskose-manager}), and how we implemented the \textit{{Packager}} for generating a Docker-based artifacts from the TOSCA specification of an application (Sect.~\ref{sec:packaging}).

\section{Managing containerised components}
\label{sec:multiple-components-per-container}
\label{sec:toskose-unit}
\subsection{Managing Multiple Components in a Container}
The first challenge to tackle for enabling an approach like ours consists in allowing multiple components to coexist within a same Docker container, with each component also being independently manageable from the other components and from the container hosting them.
Docker envisions the possibility of running multiple components in the same container\cite{docker-multi-service}, by also recommending two possible solutions for doing so. 

A na\"ive solution consists in wrapping all commands to install, configure and start the components to be hosted on a Docker container in a shell script. 
Bash job control can also be exploited to write down the shell script, \eg to delay the starting of some components, or to execute processes implementing some of their management operations.
The obtained shell script can then be executed as the main process of the container, hence meaning that the container continues to run until the shell script continues to run. 
This means that the shell script must not terminate until at least one of the components running in the container has to continue to run.
Hence, even if the na\"ive solution allows to run multiple components in the same container, their management lifecycle is still coupled to that of their hosting container.
Another example supporting this statement comes from the restarting of a component that got stopped, which requires an application administrator to force a new execution of the shell script by tearing down its hosting container and starting a new container.
The na\"ive approach indeed does not support remotely orchestrating the lifecycle of application components, in a way that is independent from the lifecycle of their hosting containers.

The other suggested solution is to exploit a process management system, like \textit{Supervisor}.
\textit{Supervisor} (\url{http://supervisord.org}) allows users to control multiple processes on Unix-like operating systems, based on a client/server model. 
A server (called \textit{supervisord}) is indeed responsible of starting sub-processes on demand, under client invocation or for handling some events (\eg for restarting a child process that crashed or exited).
\textit{supervisord} manages each spawned sub-process for the entirety of its lifetime, by also taking care of signal management, logging and configuration (including auto-starting and restarting policies).
Clients can then ask \textit{supervisord} to spawn sub-processes through a command-line interface (called \textit{supervisorctl}) or via a XML-RPC API (served by an HTTP server).
Users can also customise \textit{supervisord} by exploiting a configuration file, called \textit{supervisord.conf}. 
The latter is loaded when Supevisor starts, and it is exploited to configure \textit{supervisord}, \textit{supervisorctl} and the HTTP-served XML-RPC API.
This includes the definition of so-called \texttt{program} section, allowing to define and configure sub-processes that can be spawned on demand.

The \textit{Supervisor}-based solution is hence more suited to our needs.
Ad-hoc \texttt{program}s can be defined to allow independently executing the artifacts implementing the management operations of the components (with each \texttt{program} section devoted to a different lifecycle operation of a different component).
Such operations can then be remotely orchestrated using the XML-RPC API natively offered by \textit{Supervisor}.
Following this initial idea, we hereafter show how we exploited \textit{Supervisor} to implement the {\unit}s in our orchestration approach (Fig.~\ref{fig:big-picture}).
More precisely, we first discuss the issues characterising the management of multiple components within a same container and how \textit{Supervisor} can help solving them (Sect.~\ref{ssec:managing-processes-container}).
We then illustrate how we implement {\unit}s as \textit{Supervisor} instances (Sect.~\ref{ssec:implementing-workers}). 

\subsection{Signal Management and Zombie Reaping}
\label{ssec:managing-processes-container}
Managing multiple components in a same container inherently requires to be able to manage multiple processes in such container. 
A process indeed runs as the main process of the container (\ie the process with PID 1), and each operation to manage a component requires to spawn a corresponding sub-process, \eg executing the shell script implementing such operation.
Two subsequent, potential issues must be taken into account while managing multiple processes within the same container, \ie \textit{signal management} and \textit{zombie reaping}\cite{docker-cookbook}.

\subsubsection{Signal Management} 
Signals sent to a Docker container are forwarded to its main process, which has hence to be configured so as to allow it to decide whether and how to forward them to its sub-processes.
A striking example in this direction is the following: Suppose that a Docker container is stopped, by issuing the \texttt{docker stop} command.
The latter sends a \texttt{SIGTERM} to the main process of the container for terminating its execution\cite{docker-cli}. 
If the main process has not been configured to handle \texttt{SIGTERM}, it does not forward such signal to its child processes, which are hence not aware that the container is going to be dismissed. 
Afterwards, the Docker runtime dismisses the container by sending a \texttt{SIGKILL} signal, which results in killing uncleanly all processes running within the container. 
\texttt{SIGKILL} cannot be trapped, blocked or ignored and the processes are interrupted abnormally, possibly causing inconsistencies or data corruption\cite{advanced-linux-programming}. 


The main process of the container has hence to be configured to handle the signals it receives, by properly forwarding them to the processes running in the container.
If adopting the na\"ive solution, this drastically complicates the writing of the shell script running as the main process of the Docker container, for which we would still have the issue of not being able to remotely orchestrate the lifecycle of the components running within the container.
\textit{Supervisor} instead natively supports signal management, hence making it more suitable to our needs.

\subsubsection{Zombie Reaping} 
Another challenge while trying to manage multiple processes within the same Docker container comes from the well-know PID 1 zombie reaping problem. 
In Unix-like systems, zombie processes are processes in a terminated state, waiting for their parent to exit completely and get their descriptor removed from the process table. The descriptor of a terminate process is kept in the process table until its parent reads its exit status and remove its descriptor from the table, hence "reaping the zombie" process\cite{advanced-linux-programming}. 
Unix-based systems are typically provided with full-fledged PID 1 processes (\eg \textit{systemd} in Debian), which support routines for reaping zombie processes. 
However, the PID 1 process of a Docker container is user-defined, and typically consists in the main process of the application running within the container. 
The latter typically does not feature any routine addressing the zombie reaping problem\cite{docker-cookbook}.


Zombie processes are however harmful. 
Even if they are only consuming a little amount of memory to store their process descriptor, they keep their PID occupied.
As Unix-like systems have a finite pool of PIDs, if zombies are accumulated at a very quick rate, the pool of available PIDs can be rapidly exhausted.
This would result in preventing the spawning of other processes\cite{advanced-linux-programming}.
Given that we aim to managing multiple components within the same container, and since each call to lifecycle operation of a component results in new processes getting launched, zombie reaping has to be addressed\cite{docker-cookbook}.
However, neither the na\"ive solution nor that based on \textit{Supervisor} are supporting zombie reaping\cite{docker-multi-service}.

\subsubsection{Existing Approaches}
Several solutions are addressing signal management and zombie reaping in Docker containers, with \textit{dumb-init} (\url{https://github.com/Yelp/dumb-init}), \textit{baseimage-docker} (\url{https://github.com/phusion/baseimage-docker}) and \textit{tini} (\url{https://github.com/krallin/tini}) perhaps being the most prominent examples.
They all share the baseline idea of wrapping the main process of a Docker container with a process acting as a proxy for signals and featuring a routing addressing zombie reaping.
From developers' viewpoint, they provide a much lighter and usable solution with respect to including full-fledged init process systems (\eg \textit{sysvinit}, \textit{upstart} or \textit{systemd}) within Docker containers\cite{docker-multi-service}.

The usage of \textit{dumb-init} and \textit{tini} is analogous. 
Both must be packaged within a Docker image and they must be used as main processes (wrapping the main process of the application) when a container is spawned from such image. 
\textit{dumb-init} and \textit{tini} will then take care of {zombie reaping} in a seamless way, {by also forwarding signals to the process they wrap}. 
Differently from \textit{dumb-init}, \textit{tini} is integrated with the Docker platform (since version 1.13).
A boolean flag \texttt{{-}-init} is supported by the commands \texttt{dockerd} and \texttt{docker run}, which allows to seamlessly feature signal {forwarding} and zombie reaping in a container spawned from an existing image, backed by \textit{tini}.

Solutions like \textit{dumb-init} and \textit{tini} allow to resolve issues related to signal management and zombie reaping, but they however still lack of other essential features, \eg remote control and logging.
\textit{baseimage-docker} is a step forward in this direction, as it offers an all-in-one Docker image based on Ubuntu, featuring an init process for signal management and zombie reaping.

\subsection{Our Solution}
\label{ssec:implementing-workers}
For addressing both signal management and zombie reaping, we exploited the \textit{Supervisor}-based solution recommended by Docker\cite{docker-multi-service}, in conjunction with \textit{tini} (as the latter is already fully integrated with Docker).
More precisely, we use \textit{tini} as the main process of Docker containers, wrapping a \textit{Supervisor} instance implementing the {\unit}s of our approach.
In this way, \textit{tini} cares about zombie reaping, and \textit{Supervisor} cares about signal management, while at the same time enabling to remotely manage multiple processes within a same container.
Such an approach 
brings other valuable advantages with respect to the main competitor among other existing approaches, \ie \textit{baseimage-docker}.
Among such advantages, two are worth highligthing:
\begin{itemize}
    \item 
    \textit{baseimage-docker} is coming only with a given distribution of Ubuntu, while \textit{Supervisor} and \textit{tini} work with any operating system distribution featured by a Docker container.
    This hence makes our approach applicable to a wider set of scenaria.
    \item
    \texttt{baseimage-docker} only support SSH to remotely access the internals of a container.
    \textit{Supervisor} instead exposes a customisable XML-RPC API on top of a HTTP server. 
    The API exposes methods for managing the lifecycle of both \texttt{supervisord} and its child processes, and it can be customised by exploiting an external INI-style configuration file.
    In particular, it is possible to define \texttt{program} sections, which result in offering remotely accessible methods that can be invoked on demand.
    The latter acts as an enabler for our orchestration approach, as the management operations associated with the components of an application can be implemented as \textit{Supervisor} \texttt{programs}.
\end{itemize}

{\unit}s are implemented by packaging a (\textit{tini}-wrapped) standalone instance of \textit{Supervisor} in each container running one or more application components. 
It runs as the main process of the container, and it is configured to allow executing (on demand) the artifacts implementing the management operations of the components hosted by the container.
The latter is obtained by providing the \textit{Supervisor} instance with a configuration file containing a different \texttt{program} section for each management operation supported by the components hosted by the container, hence configuring the XML-RPC API of the \textit{Supervisor} instance to feature a remotely accessible operation for each management operation of hosted components.
The configuration file is automatically generated from the TOSCA specification of an application, and it is automatically packaged within each container of an application together with \textit{Supervisor} (Sect.~\ref{sec:packaging}).

To enable the packaging of a standalone instance of \textit{Supervisor}, we developed \toskoseunit (\url{https://github.com/di-unipi-socc/toskose-unit}).
\toskoseunit is a Docker image bundling a standalone instance of \textit{Supervisor}, which is publicly available on the Docker Hub (\url{https://hub.docker.com/r/diunipisocc/toskose-unit}).
Its purpose is to allow including a suitably configured instance of \textit{Supervisor} in any Docker image of the containers forming the infrastructure of an application, which can be obtained by means of multi-stage Docker builds.

%

While developing \toskoseunit, we had to address an issue inherent to the usage of \textit{Supervisor} within a Docker container.
\textit{Supervisor} is a Python application, hence requiring Python runtime to be available in the container running it. 
However, installing a Python interpreter in a Docker image can generate conflicts, if the Docker image is already featuring some Python interpreter.
To address this issue, we exploited the PyInstaller ``freezing'' tool.
PyInstaller ``freezes'' an existing Python program by creating a single-file executable that contains the application code and the Python interpreter to run it.
In this way, we ``freezed'' \textit{Supervisor} and created a bundle not needing any Python interpreter or module to be installed in the Docker containers running it, hence avoiding the risk for conflicts.

\section{Orchestrating multi-component applications}
\label{sec:orchestration}
\label{sec:toskose-manager}
The second challenge to tackle consists in allowing to seamlessly manage the lifecycle of the components forming a TOSCA application, by enabling the remote invocation of their management operations, and by running them on top of existing Docker container orchestrators.
Of course, the latter is because we wish to avoid reinventing the wheel, \ie instead of re-designing a container orchestrator from scratch, we wish to piggyback on top of existing, production-ready orchestrators, also for allowing to reuse the capabilities they feature.

Existing container orchestrators already allow deploying and managing containers.
For instance, given a specification of the containers to run and of their configuration, both Docker Swarm and Kubernetes can deploy such containers on a cluster of hosts by also configuring them as indicated.
Docker Swarm and Kubernetes then proceed by orchestrating deployed containers, \eg by applying them load balancing and scaling policies, for recovering failed instances, and to manage overlay networks or provisioning resources\cite{docker-cookbook}.

At the same time, Docker containers are treated as "black-boxes" by existing container orchestrators, as the minimal entity that can be orchestrated is the container itself\cite{tosker}.  
Our objective is to go a step forward, by enabling a component-aware orchestration of containerised applications, \ie we aim at allowing to orchestrate the management of components running within the same container.
To this end, we introduced {\unit}s (\ie suitably configured Supervisor instances --- Sect.~\ref{sec:toskose-unit}) in each container running some component, so as to offer an XML-RPC API allowing to remotely invoke the operations for managing the lifecycle of the components it hosts.
The next step is hence to find a suitable implementation of the {\manager} in our orchestration approach (Fig.~\ref{fig:big-picture}), \ie to introduce a containerised component in an application, which allows coordinating the management of each of its components by suitably interacting with the corresponding {\unit}.

\begin{figure}
  \centering
  \includegraphics[scale=\figscale]{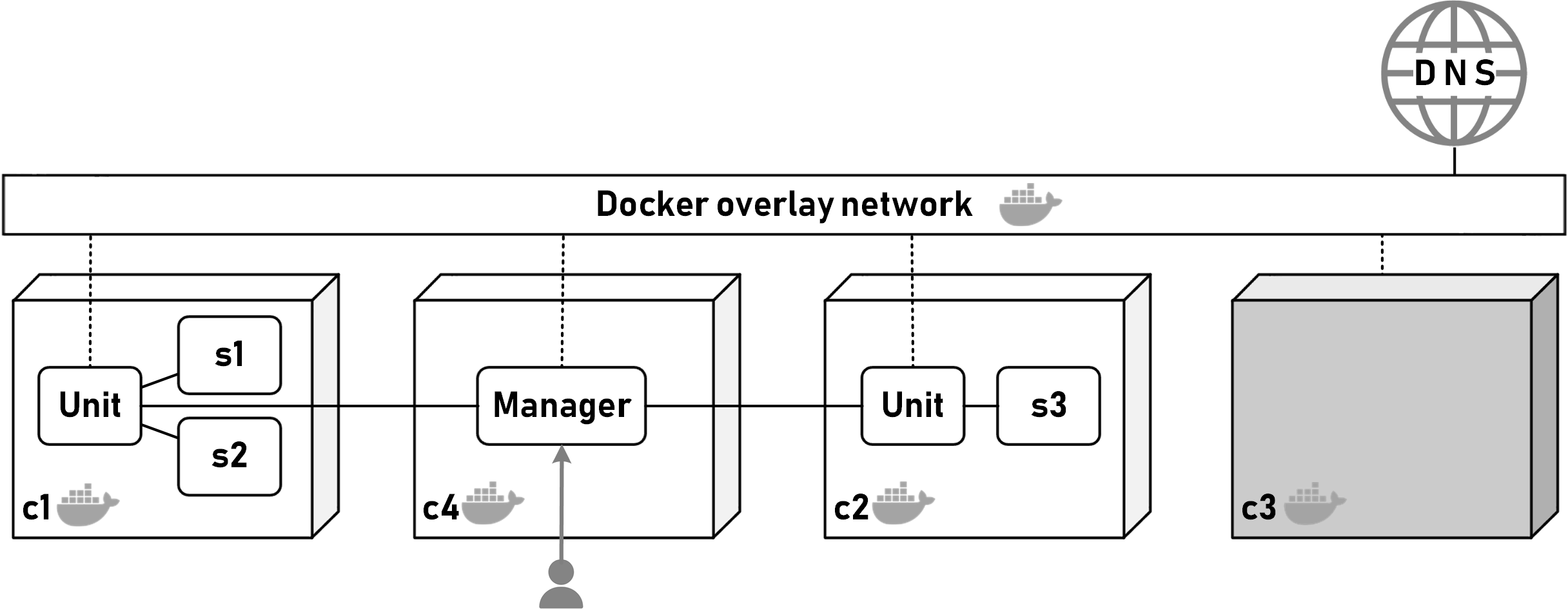}
  \caption{Interactions among {\manager} and {\unit}s for managing the lifecycle of containerised application components.}
  \label{fig:toskose-manager-orchestration}
\end{figure}

Our baseline idea for doing so is illustrated in Fig.~\ref{fig:toskose-manager-orchestration}.
Whenever the user wishes to execute a management operation on a component of her application, she invokes the {\manager} asking it to execute such operation.
The {\manager} then forwards the request to the {\unit} managing such component (\ie it invokes the XML-RPC API of the Supervisor instance running within the container hosting the component).
The interaction between the {\manager} and the {\unit} occurs through a Docker overlay network, and as soon as the {\unit} receives the request, it executes the required management operation (\ie it runs the corresponding \texttt{program} section of the \textit{Supervisor} configuration file).
We hereafter illustrate a possible realisation of such an idea, given by the \toskosemanager.

\subsection{The Architecture of the \toskosemanager}
\label{ssec:toskose-manager-architecture}
The \toskosemanager realises the {\manager} in our orchestration approach (Fig.~\ref{fig:big-picture}), hence being responsible of coordinating {\unit}s to allow remotely managing the containerised components of an application, based on its TOSCA specification.
{The \toskosemanager is intended to be included in an application as an additional containerised component, as this will allow managing it with Docker-based container orchestrators, together with all other containers forming an application.%
}

Given that we are piggybacking on top of existing container orchestrators, we can exploit their capabilities for setting up the overlay network where the containers of an application will run (both for single-host and multi-host settings).
This also means that, by properly setting network aliases, containers can intercommunicate by simply exploiting their hostnames on the overlay network, which will be automatically resolved by the network DNS.
%
In addition, Docker-based container orchestrators allow running different applications in different virtual private networks.
This has two main advantages for our purposes, namely (i)~it permits securing the interactions among the components of an application, including the \toskosemanager, and (ii)~even if multiple instances of \toskosemanager run on the same overlay network, they will not interfere one another.

With the above setting, \toskosemanager only requires to know which container is hosting which components (to be able to interact with the {\unit} running in the same container) and the network aliases associated to the containers of an application.
Such information is provided to \toskosemanager by feeding it with the TOSCA application specification (from which it can retrieve the relations among components and containers) and an additional configuration file containing the network aliases associated to the container.
The configuration file can be automatically generated, and both files are automatically injected to the \toskosemanager by the \toskosepackager (Sect.~\ref{sec:toskose}).

\begin{figure}
  \centering
  \includegraphics[scale=\figscale]{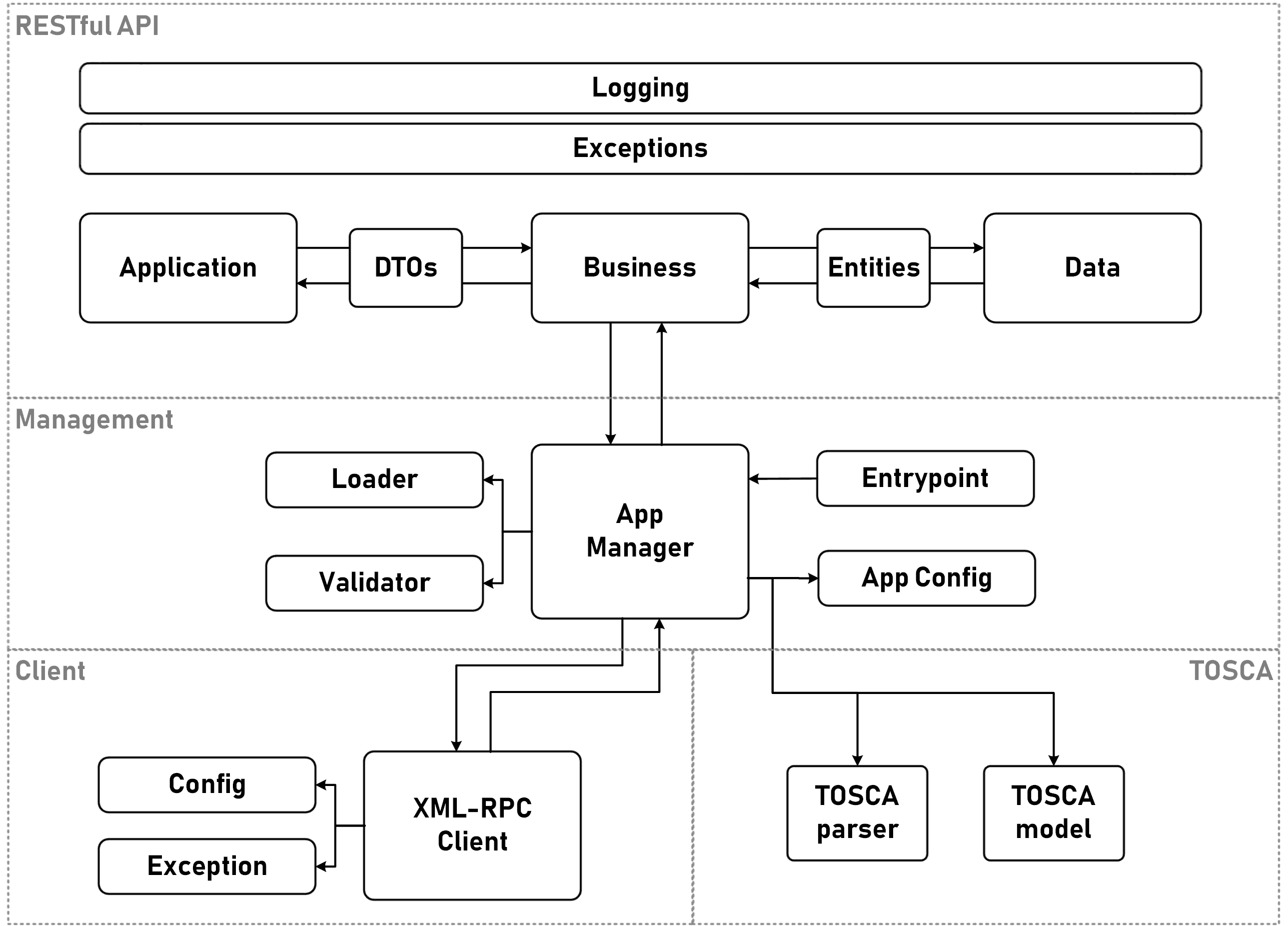}
  \caption{The architecture of the \toskosemanager.}
  \label{fig:toskose-manager-architecture}
\end{figure}

To concretely implement the baseline idea shown in Fig.~\ref{fig:toskose-manager-orchestration}, the \toskosemanager must hence feature 
(i)~a remotely accessible API for allowing an application administrator to invoke the operation to manage the components of her application, 
(ii)~a core processing module for identifying the {\unit}s where to forward requests, and 
(iii)~a client for the XML-RPC API offered by the {\unit}, to concretely forward requests.
It must also be capable of 
(iv)~processing TOSCA application specifications.
Fig.~\ref{fig:toskose-manager-architecture} shows the architecture of \toskosemanager, designed to feature (i-iv), as well as to comply with the separation of concerns design principles to make it modular and extensible\cite{software-engineering-book}.

\smallskip \noindent
\textbf{RESTful API.}
A \textit{RESTful API} allows the application administrator to invoke the execution of a management operation on a component of her application.
The API is designed by following the Web API design paradigm\cite{google-api-design}, by partitioning it in three main logical layers, \ie \textit{Application}, \textit{Business} and \textit{Data}.
The \textit{Application} layer contains the controllers for translating HTTP incoming requests and outgoing responses, and for encoding and decoding their payloads, by also validating them.
The \textit{Business} layer is where the business logic of the API resides, with business rules and workflows defined to suitably interact with the \textit{Data} layer and with the core of \toskosemanager.
The \textit{Data} layer provides a storage of all information needed to orchestrate an application (\ie component and container names, network aliases, etc.).

To enhance data encapsulation in inter-layer communications, \textit{DTOs} and \textit{Entities} are defined. 
These are used in the communication between the \textit{Application} and \textit{Business} layers, and between the \textit{Business} and \textit{Data} layers, respectively.
Two other standalone modules are used for logging and error handling of the API, \ie \textit{Logging} and \textit{Exceptions}, which are organised as cross-cutting concerns.
    
\smallskip \noindent
\textbf{Management.}
The \textit{Management} area contains the core module of \toskosemanager, \ie \textit{{App Manager}}.
The latter acts as a proxy between the \textit{RESTful API} and the \textit{Client} modules.
It indeed receives requests from the API (\eg executing a management operation on a component, or getting the status of all components), and it suitably interacts with the \textit{Client} modules so that they interact and coordinate {\unit}s to carry out the requests.
    
The \textit{Management} is also responsible of configuring the environment for allowing \textit{{App Manager}} to run.
It indeed contains the \textit{App Config} and \textit{Entrypoint} modules, 
{which configure the runtime environment and run a web server used to host the \textit{{App Manager} } (with the web server also being the main process run in the container of the \toskosemanager)}.
It also contains the the \textit{Loader} and \textit{Validator} modules, used for loading and validating the TOSCA application specification and the Toskose configuration file, which contain the information needed to orchestrate the specified application. 
    
\smallskip \noindent
\textbf{Client.}
The \textit{Client} is responsible of communicating with the {\unit}s on the containers hosting the components of an application. 
It hence contains a \textit{XML-RPC Client} allowing to invoke the XML-RPC API offered by the \textit{Supervisor} instances implementing the {\unit}s.
Whenever the \textit{XML-RPC Client} is required by the \textit{{App Manager}} to invoke an operation offered by a {\unit}, it builds and sends a HTTP request to the API of such {\unit}, which payload is structured according to the XML structure expected by the API of \textit{Supervisor}.
The \textit{XML-RPC Client} is returned XML data representing the outcome of its request from the {\unit}, and it communicates the \textit{{App Manager}} such an outcome.
    
For enforcing fine-grained failure management, and in accordance to separation of concerns design principles, error handling is kept separate from the rest of the application\cite{software-engineering-book}.
Errors are indeed handled by the \textit{Exceptions} module, which is also part of the \textit{Client} area.

\smallskip \noindent
\textbf{TOSCA.}
The \textit{TOSCA} area is responsible of the processing of TOSCA application specifications.
It indeed features two modules, \ie \textit{TOSCA parser} and \textit{TOSCA modelling}, intended to allow parsing TOSCA application specifications and to build an in-memory representation of specified applications.

\subsection{A Prototype Implementation of the \toskosemanager}
An open-source prototype implementation of the \toskosemanager is publicly available on GitHub (\url{https://github.com/di-unipi-socc/toskose-manager}), and it is also shipped as a Docker image publicly available on the Docker Hub (\url{https://hub.docker.com/r/diunipisocc/toskose-manager}).
The prototype of \toskosemanager is written in Python (v3.7.1), and we hereafter detail its implementation.
More precisely, given that the \textit{TOSCA} modules have been obtained by suitably extending the OpenStack TOSCA parser (\url{https://github.com/openstack/tosca-parser}), we shall focus on the implementation of the \textit{RESTful API} and of the \textit{Management} and \textit{Client} areas.

\begin{figure}
  \centering
  \includegraphics[scale=\figscale]{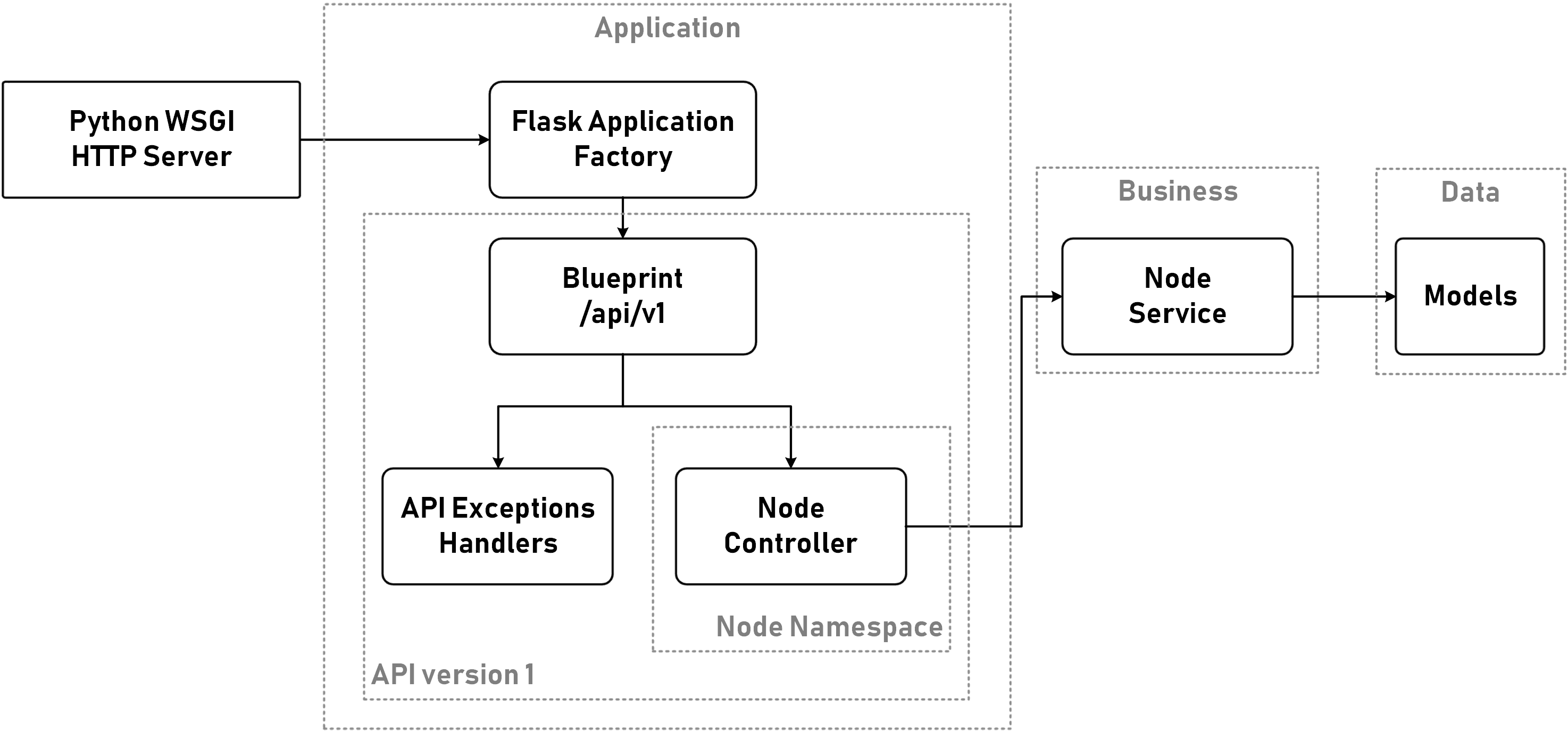}
  \caption{Architecture of the prototype implementation of the \textit{RESTful API} of \toskosemanager.}
  \label{fig:toskose-manager-api-design}
\end{figure}

\smallskip \noindent 
\textbf{RESTful API.} 
Fig.~\ref{fig:toskose-manager-api-design} illustrates the architecture of the \textit{RESTful API} featured by the prototype implementation of \toskosemanager.
The topmost component is a \textit{Python WSGI HTTP Server}, where WSGI stands for "Web Server Gateway Interface" (which is a specification describing how a web server communicates with the web applications it hosts, and how they can be chained together to process a request).
The HTTP Server is powered by Gunicorn (\url{https://gunicorn.org}), it implements the WSGI interface, and it permits running the web application implementing the API to be offered.

The \textit{Application} layer of the API is then implemented by exploiting the Flask Python framework (\url{https://palletsprojects.com/p/flask}), extended with Flask-RESTPlus (\url{https://flask-restplus.readthedocs.io}).
The latter has been included as it provides a collection of Python decorators and tools for quickly building RESTful APIs and exposing their documentation using the Swagger UI (\url{https://swagger.io}).
In addition, Flask-RESTPluse enforces modularity of built APIs, hence making the current implementation of the \textit{Application} layer of the \textit{RESTful API} extensible for further developments.

The above setting has been obtained by suitably configuring the \textit{Flask Application Factory}, which acts as the entrypoint of an application, by initialising the Flask environment where the application will run.
This includes mounting extensions (such as Flask-RESTplus), as well as registering blueprints (\ie logical groups partitioning the modules of an application based on the concerns they relate to, to enforce separation of concerns\cite{software-engineering-book}).
For instructing the \textit{Flask Application Factory} to loading the current prototype of the API, we hence developed a \textit{Blueprint /api/v1}.
Notice that adding a different version of the API, or running different versions simultaneously, simply require to change or add another blueprint among those registered.

\begin{figure}
    \centering
    \includegraphics[width=.99\textwidth]{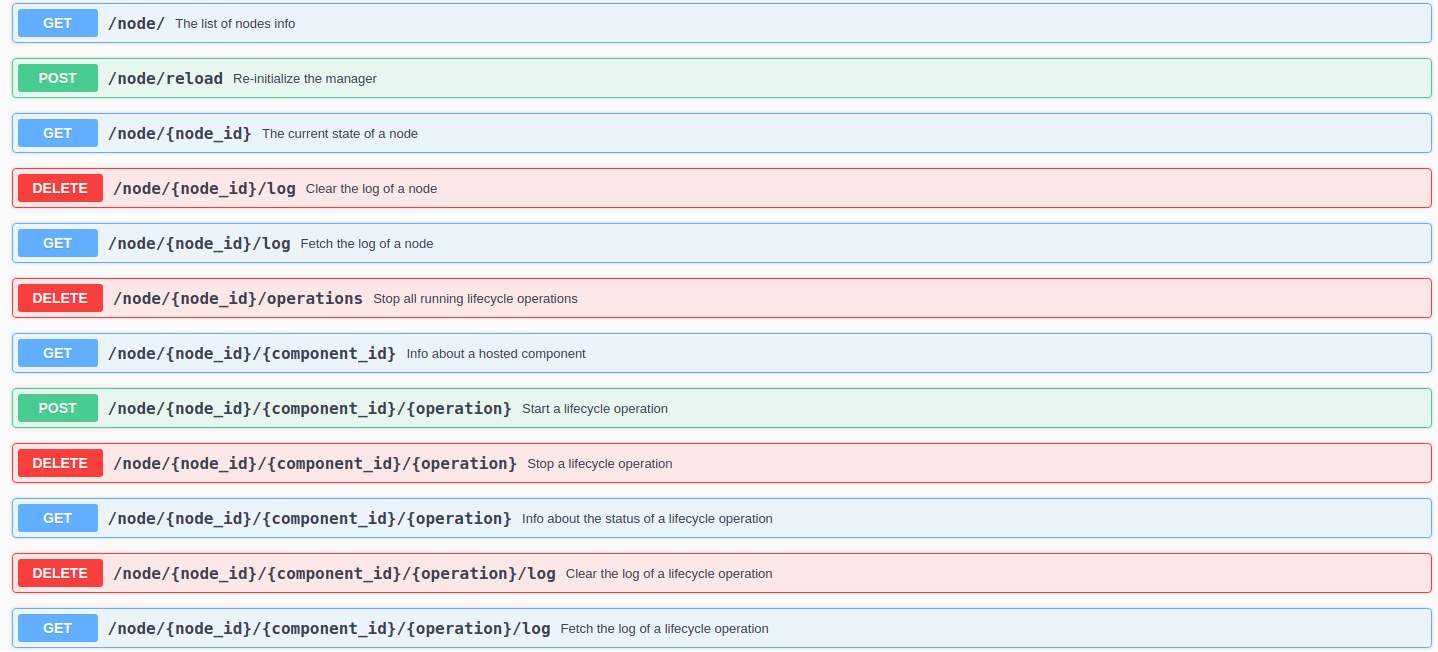}
    \caption{Snapshot of the HTTP methods offered by a running instance of the \textit{RESTful API} of \toskosemanager, obtained from the Swagger UI featured by Flask RESTPlus.}
    \label{fig:restful-api-methods}
\end{figure}

In addition, by exploiting the Flask RESTPlus extension, we logically organised the Blueprint using so-called "namespaces". 
A \textit{Node Namespace} is indeed used, which is exploited to mark the \textit{Node Controller} related to the resource \texttt{/node}.
The latter is the root resource of \textit{RESTful API}, and the \textit{Node Controller} has been implemented so as to offer all methods shown in Fig.~\ref{fig:restful-api-methods}
The \textit{Node Controller} is hence responsible of accepting incoming requests, which it decodes and validates by checking their payload against an expected schema.
Non-valid requests are refused, while valid requests are passed to the business layer for processing.
Such a forwarding involves exchanging complex structured data, which is done by exploiting DTOs for enforcing data encapsulation.

The \textit{Business} layer is implemented by the \textit{Node Service}, which business rules allow to process incoming requests.
Intuitively, it processes each request by retrieving information on involved application components, aggregating such data in the form of DTOs, and passing the request and retrieved data to the core of \toskosemanager.
In addition, the \textit{Node Service} fetches and manipulates the logs of the \textit{RESTful API}.

\smallskip \noindent 
\textbf{Management.}
The \textit{{App Manager}} module in the \textit{Management} area (Fig.~\ref{fig:toskose-manager-architecture}) is the core module for managing multi-component application with \toskose.
Such a module maintains an in-memory representation of the managed application built from its TOSCA specification and enriched according to the Toskose configuration file, both loaded and validated during the initialization of the module. 
The TOSCA parsing is done by exploiting another Python module (\ie \textit{TOSCAParser}), which implementation is essentially wrapping the OpenStack TOSCA Parser library.
Logging and error handling are also configured during the initialization of \textit{{App Manager}}, by exploting the standalone \textit{Loader} module.

After the initialization, the \textit{{App Manager}} starts waiting for incoming requests for the (\textit{Node Service} of the) \textit{RESTful API}.
Upon the receiving of a request, it instructs the \textit{XML-RPC Client} to contact the {\unit} managing the component involved by the request.

\smallskip \noindent 
\textbf{Client.}
The \textit{Client} area is implemented following the Factory method pattern, with the \textit{{App Manager}} delegating to the \textit{Client Factory} the decision to determine the concrete implementation of the client that must be used to interact with a {\unit}.
Currently, the only available implementation is that for the XML-RPC API of \textit{Supervisor}, which is obtained by exploiting the Python built-in \texttt{xmlrpc.client}.
We anyway decided to implement clients using the Factory method pattern to allow \toskosemanager to be extended to support multiple interaction protocols, in view of further developments.

\section{Generating deployable artifacts}
\label{sec:packaging}
The last brick needed for enabling our orchestration approach is the {\packager}, \ie a solution that, given the TOSCA specification of an application, automatically generates the deployable artifacts needed to actually enact its deployment on top of existing, production-ready container orchestrators. 
The deployable artifacts must be such that {\unit}s (\ie suitably configured Supervisor instances) are included within the containers hosting application components, and that a containerised \toskosemanager is added to the application.
The {\unit}s and the \toskosemanager are indeed needed to enable a component-aware orchestration on top of the targeted container orchestrator.
We hereafter present our solution for doing so, by first illustrating a solution for automating the generation of deployable artifacts, and by then presenting its prototype implementation, \ie \toskosepackager.

\subsection{Automating the Generation of Deployable Artifacts}
\begin{figure}
  \centering
  \includegraphics[scale=0.5]{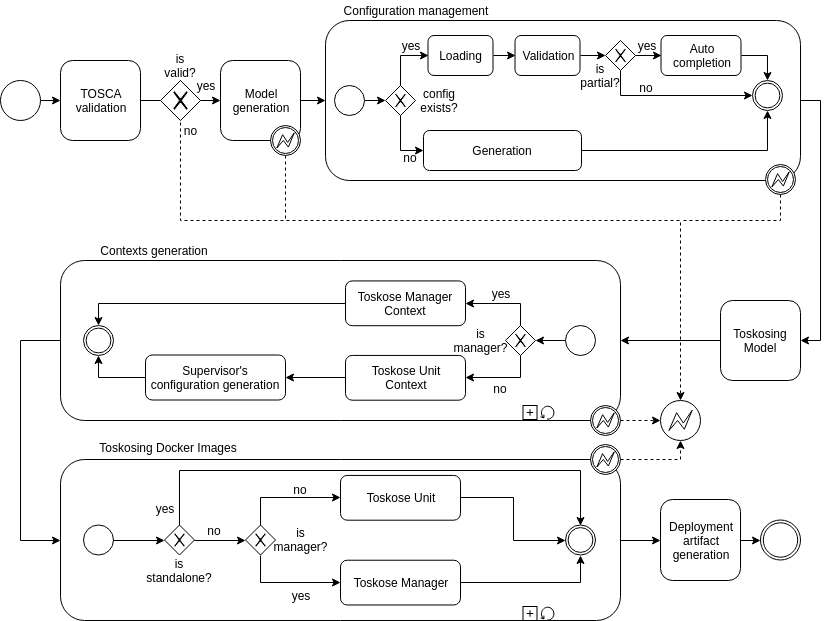}
  \caption{Workflow for automatically generating deployable artifacts, depicted according to the BPMN graphical notation\cite{bpmn}.}
  \label{fig:toskose-pipeline}
\end{figure}

Fig.~\ref{fig:toskose-pipeline} illustrates a workflow that, given the TOSCA specification of an application, automatically generates the deployable artifacts for enabling a component-aware orchestration of multi-component applications on top of an existing container orchestrator.
The workflow begins with the activities \textit{TOSCA validation} and \textit{Model generation}, which validate the TOSCA application specification and generate an in-memory representation of the specified application, respectively.
If the given TOSCA specification is not valid, or if the \textit{Model generation} fails, the workflow exits by returning error information.

The workflow then proceeds by dealing with the \textit{Configuration management}, \ie with the file indicating the configuration of the containers forming the application, including the network aliases and ports for reaching the components they host.   
If such a configuration file is missing or partial, the workflow automatically sets the missing fields to default values.
The obtained configuration file is then used by the \textit{Toskosing model} activity, which enriches the in-memory application representation with the information contained in the configuration file, as well as by adding the additional container packaging the \toskosemanager.

The subsequent phase (\textit{Contexts generation}) is repeated for each container of the application hosting some component (either being some original application component or the newly introduced \toskosemanager).
For each of such container, a Docker build context is prepared, by setting up a folder containing all files needed to build a Docker image.
If a container is hosting application components, a so-called \textit{Toskose Unit Context} is generated.
The latter contains all the artifacts needed for deploying and managing the application components hosted on the container, \ie the artifacts implementing an application component and those implementing its management operations.
An automatically generated \textit{Supervisor} configuration file is also included within the build context, which will then be used to configure the \textit{Supervisor} instance implementing the {\unit} on the container (\eg for ensuring that its XML-RPC API will offer methods for remotely invoking the management operations of the hosted components).
If the processed container is instead that hosting the \toskosemanager, the build context only contains the TOSCA specification and the Toskose configuration file of the application under processing.

The workflow then proceeds by \textit{Toskosing Docker images}, \ie preparing the Docker images of each containers of the application.
If the container is a standalone container, the original image is kept. 
If instead the container is hosting some application component, two different activities are enacted, depending on whether the hosted components are original application components or the \toskosemanager.
In the former case, a new Docker image is built by combining the corresponding build context and the \textit{Supervisor} bundle fetched from the \toskoseunit Docker image, by means of a multi-stage Dockerfile.
In the latter case, the build context is instead combined with the Docker image packaging the \toskosemanager, still by means of a multi-stage Dockerfile.

The deployable artifact generation process then completes with the \textit{Deployment artifact generation} activity, which essentially takes the newly generated images of Docker containers (hereafter called \textit{toskosed} images, for brevity) and combines them in a multi-container application deployment, \eg a Docker Compose file. 
The \textit{toskosed} images are configured in accordance with the in-memory application representation, so as to allow them to properly intercommunicate (both for running the application business and for allowing the \toskosemanager to interact with the \textit{Supevisor} instances implementing the {\unit}s).

\subsection{The Architecture of the \toskosepackager}
\begin{figure}
  \centering
  \includegraphics[scale=\figscale]{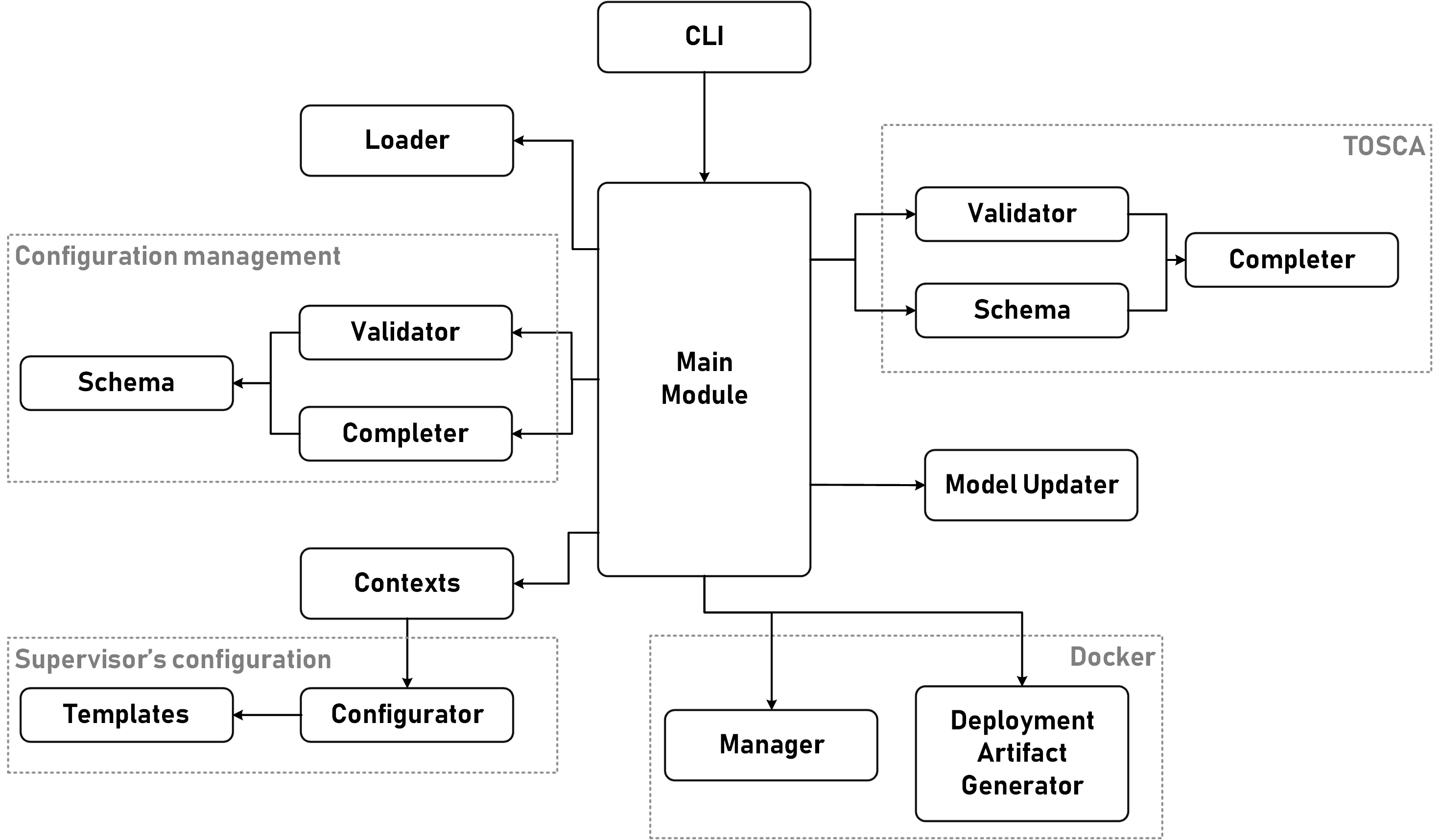}
  \caption{The architecture of the \toskosepackager.}
  \label{fig:toskose-architecture}
\end{figure}

Fig.~\ref{fig:toskose-architecture} illustrates the architecture of \toskosepackager, our solution for generating the artifacts enabling the component-aware orchestration of an application on existing Docker-based container orchestrators, based on the workflow in Fig.~\ref{fig:toskose-pipeline}.
Still pursuing the aim of obtaining a modular and extensible solution, the architecture is designed in accordance with the separation of concerns design principles\cite{software-engineering-book}.
The architecture is indeed partitioned by devoting different sets of modules to different stages of the workflow.

A \textit{CLI} module allows users to provide the \toskosepackager with the necessary input for starting the process for automating the generation of deployable artifacts, \ie a TOSCA application specification and (optionally) a Toskose configuration file.
The input is then passed to the \textit{Main module}, which is in charge of coordinating the modules forming the architecture of \toskosepackager to suitably execute the activities of the workflow.
It then first processes the input, by invoking the \textit{Loader} module for actually importing the input files, and the modules in the \textit{TOSCA} area for validating the TOSCA application specification and generating an in-memory representation the application.

The \textit{Main Module} then interacts with the modules in the \textit{Configuration management} area.
The \textit{Validator} module allows the \textit{Main module} to validate the input Toskose configuration file, if any.
The \textit{Completer} module instead allows the \textit{Main module} to complete the Toskose configuration file with default configurations, or to generate it from scratch if no configuration file is provided.
Both the \textit{Validator} and the \textit{Completer} relies on a \textit{Schema} module, indicating how the Toskose configuration file has to be structured, and which default values to employ for missing configurations.
Once the configuration is ready, the \textit{Model Updater} is used by the \textit{Main Module} to enrich the in-memory application representation with the configuration information. 

The \textit{Main Module} continues by interacting with the \textit{Contexts} module, which creates the build contexts for each container hosting some component, including that hosting the \toskosemanager.
For the containers hosting original application components, \textit{Contexts} interacts with the \textit{Configurator}, which allows to create the file for configuring the \textit{Supervisor} instances implementing the {\unit}s, based on existing \textit{Templates}.

Finally, the \textit{Main Module} interacts with the modules in the \textit{Docker} area.
It first requires to the \textit{Manager} to build the Docker images for the containers forming the application, which the \textit{Manager} carries out by relying on the build features of the Docker Engine.
The \textit{Main Module} then requires the generation of the final \textit{Deployment artifact} to the \textit{Deployment Artifact Generator}.


\subsection{A Prototype Implementation of the \toskosepackager}
An open-source prototype implementation of the \toskosepackager is publicly available on GitHub (\url{https://github.com/di-unipi-socc/toskose-packager}).
The prototype is written in Python (v3.6) and it has been released on the Python Package Index (\url{https://pypi.org/project/toskose}), to allow installing it with the command \texttt{pip install toskose}.

The prototype provides a command-line interface that takes as input a Cloud Service ARchive and (optionally) a Toskose configuration file, and it returns a Docker Compose artifact.
The latter allows to deploy the application on Docker-based container orchestrators, with the latter being used to orchestrate the components of the application, while the orchestration of the components hosted on the containers being enabled by the RESTful API of \toskosemanager (Sect.~\ref{sec:toskose-manager}).
Currently, the generated Docker Compose artifact is tested and fully working on Docker Swarm and on Kubernetes, with the latter requiring to first run Kompose (\url{https://kompose.io}) or Compose Object (\url{https://github.com/docker/compose-on-kubernetes}) to actually enact the deployment of the application.

We hereafter detail our prototype implementation of the \toskosepackager, by showing how each component of the architecture in Fig.~\ref{fig:toskose-architecture} has been implemented.

\smallskip \noindent
\textbf{CLI.}
The command-line interface (\textit{CLI}) 
offers the following interface:
\begin{Verbatim}
  toskose [OPTIONS] CSAR_PATH [CONFIG_PATH]
\end{Verbatim}
where the optional argument \texttt{OPTIONS} is a list of options for customising the run of \texttt{toskose}. In particular, \texttt{-o} and \texttt{{-}-output-path} \texttt{PATH} allow to specify the path where to place the output deployment artifacts, \texttt{-p} and \texttt{{-}-enable-push} activate the automatic pushing of \textit{toskosed} images on a Docker registry, \texttt{{-}-docker-url} \texttt{URL} allows to define a custom entrypoint for the Docker Engine API, \texttt{-q} and \texttt{{-}-quiet} reduce the output information messages, and \texttt{{-}-debug} activates the debug mode.

The other arguments instead indicate the paths to the input files for the \toskosepackager.
More precisely, \texttt{CSAR\_PATH} indicates the path to a Cloud Service ARchive (CSAR), containing the TOSCA specification of an application and the artifacts realising the application and its management operations.
The optional argument \texttt{CONFIG\_PATH} instead provides the path to a Toskose configuration file.

\smallskip \noindent
\textbf{Main Module.}
The \textit{Main Module} is implemented as a Python module that is invoked by the command-line interface, which passes it the paths to the input files to be processed, and the processing options.
It then starts coordinating the other modules of our prototype implementation of \toskosepackager to carry out the activities of the workflow in  Fig.~\ref{fig:toskose-pipeline}, in the given order. 

The \textit{Main Module} also generates to temporary directories using the Python \texttt{tempfile} library. 
A directory is used for storing the content of the (unpacked) CSAR archive, while the other one is used for storing the Docker build contexts. 
Both the directories are stored under \texttt{/tmp} and they are removed once the \textit{Main Module} reaches the end of the workflow.

\smallskip \noindent
\textbf{TOSCA.}
The implementation of the \textit{Validator} provides the necessary for checking whether the input CSAR archive complies with the TOSCA standard\cite{tosca}. 
It indeed allows to check whether the extension of the archive complies with admitted ones (\ie \texttt{.csar} or \texttt{.zip}), as well as the directories composing the archive are organised as indicated by the TOSCA standard.

The \textit{Parser} and \textit{TOSCA Model} {instead allow parsing the YAML file defining the TOSCA specification of an application and building an in-memory representation of the application.
They are implemented by extending the OpenStack TOSCA parser (\url{https://github.com/openstack/tosca-parser}), in order to allow processing the TOSCA-based representation given by TosKer\cite{tosker}.}

\smallskip \noindent
\textbf{Configuration.}
Toskose configuration files are specified in YAML, and they are structured in two YAML objects, \ie \texttt{nodes} and \texttt{manager}.
The object \texttt{nodes} is devoted to the configuration of the containers hosting application components, with one nested YAML object for each of such containers.
Each nested object is given the same name as that of the container in the TOSCA application specification, and it allows to specify the \texttt{alias} associated to the container and the \texttt{port} where the XML-RPC API of the \textit{Supervisor} instance implementing a \unit is offered.
These are then used by the used by the \toskosemanager for communicating with the \unit managing the components running in the container.

The object \texttt{manager} instead allows to provide the configuration information for the Docker container running the \toskosemanager.
It indeed allows specifying its \texttt{alias} on the Docker network, as well as the \texttt{port}, \texttt{username} and \texttt{password} for accessing the \textit{RESTful API} of the \toskosemanager

A Toskose configuration file can be optionally provided to our prototype implementation of the \toskosepackager. 
If provided, it is validated against the above illustrated schema by the implementation of the \textit{Validator}.
If something is missing, or if no configuration file is given, an auto-completion routine implementing the \textit{Completer} allows to fill it with default values.

\smallskip \noindent
\textbf{Model Updater.}
The implementation of the \textit{Model Updater} adds the information contained in the Toskose configuration file to the in-memory representation of the processed application.
More precisely, it first extends the application representation by setting environment variables to be defined in each container hosting a component, in such a way that the instance of \textit{Supervisor} it runs is configured to listen on the indicated \texttt{port}. 
It also extends the representation of each container by setting properties needed for naming and tagging the correspondingly generated \textit{toskosed} image (and optionally pushing it to a Docker registry).

The implementation of the \textit{Model Updater} also includes an additional container to the in-memory application representation, devoted to hosting the \toskosemanager.
Such a container is then configured according to the specified configuration information, in a way similar to that described above.
The choice of injecting the addition container during this activity simplifies the subsequent steps, which have to process only the in-memory application representation instead of fetching information from different sources.

\smallskip \noindent
\textbf{Contexts generation.}
The \textit{Contexts} module of our prototype implementation of the \toskosepackager fills a temporary folder devoted to Docker build contexts with a \textit{Toskose Manager Context} for the container hosting the \toskosemanager and a \textit{Toskose Unit Context} for each container hosting some original application component.
%
%
The \textit{Toskose Manager Context} is devoted to storing the Toskose configuration file and the TOSCA specification of the application under processing, which are needed by the \toskosemanager for enacting the management of the application.

\begin{figure}
  \centering
  \includegraphics[scale=\figscale]{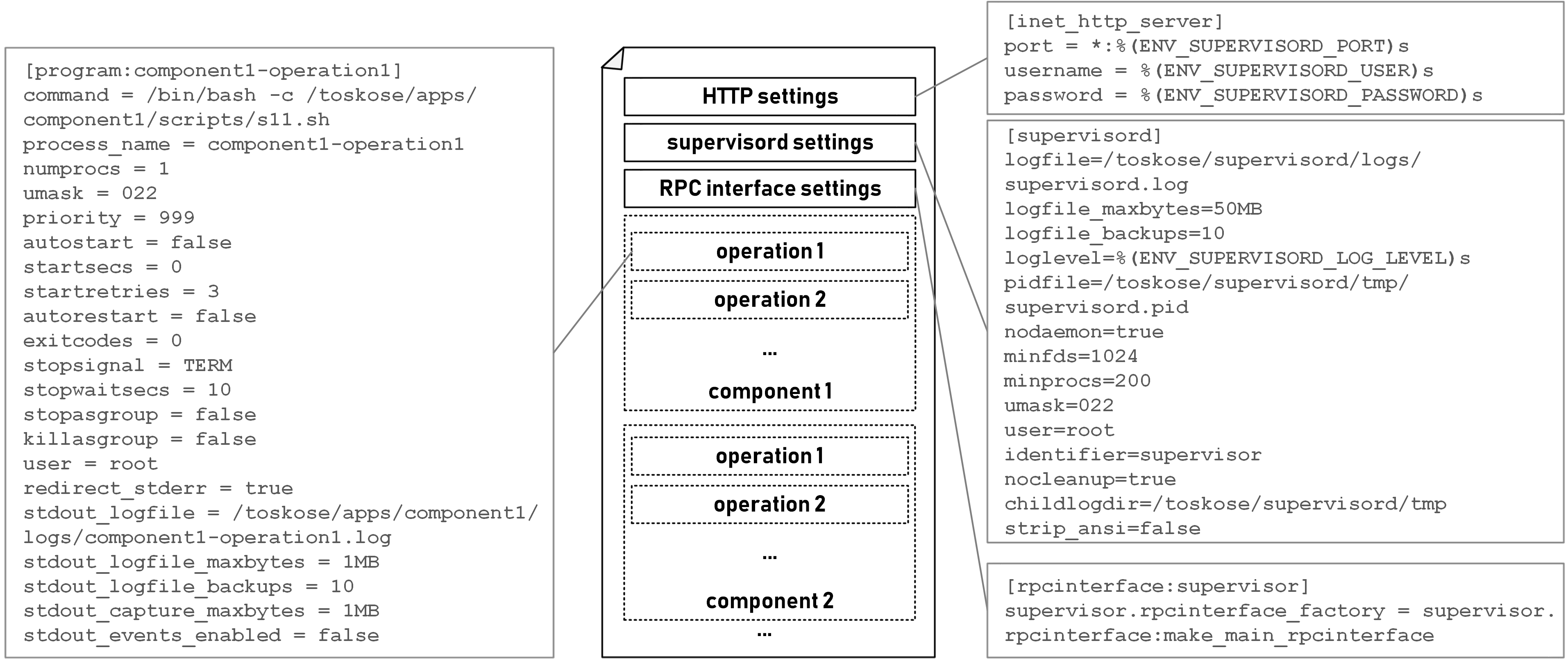}
  \caption{Template of the configuration of \textit{Supervisor} injected in Docker containers to implement {\unit}s.}
  \label{fig:toskose-supervisor-configuration-schema}
\end{figure}

Each \textit{Toskose Unit Context} is assigned the name of the corresponding container.
It is devoted to contain the artifacts realising the components hosted on the container, as well as the scripts implementing their management operations.
The folder also contains a file \texttt{supervisor.conf}, providing all configuration information needed by the \textit{Supervisor} instance running on the container for suitably implementing a {\unit}.
The file \texttt{supervisor.conf} is generated by automatically extending a base template, which schema is in Fig.~\ref{fig:toskose-supervisor-configuration-schema}.
The \textit{HTTP settings} section configures the HTTP server used for offering the XML-RPC API of \textit{Supervisor} by expanding the enviroment variables defining its configuration in the in-memory application representation.
The \textit{supervisord settings} section configures \textit{supervisord} (the main process of \textit{Supervisor}) with some default configurations, except for the logging level, which is fetched from the information stored in the in-memory application representation.
The \textit{RPC interface settings} section initialises the XML-RPC API of the \textit{Supervisor} instance to run.
%
The file is completed by including \texttt{program} sections allowing to remotely invoke the management operations of the application component hosted on the container.

\smallskip \noindent
\textbf{Docker.}
The \textit{toskosed} Docker images are built by the \textit{Manager} module.
The latter is implemented by exploiting the \texttt{docker-py} Python library (\url{https://docker-py.readthedocs.io}), which allows to interact with the API of the Docker Engine installed on a host, also for building Docker images.
For each container to be \textit{toskosed} (including the \toskosemanager), the \textit{Manager} implementation proceeds as follows.
Firstly, it pulls the \texttt{toskose-unit} or \texttt{toskose-manager} Docker image from the Docker Hub and the Docker image associated with the container in the in-memory application from the specified registry, if they are not already available locally.
It then instructs the Docker Engine to build the \textit{toskosed} image, by passing it the application context, the pulled images and a multi-stage Dockerfile.
If explicitly required by the user, the \textit{Manager} module also proceeds with pushing \textit{toskosed} images to a Docker registry.

The implementation of the \textit{Deployment Artifact Generator} completes the workflow, by generating the Docker Compose file from the in-memory application representation.
The containers of the application (including standalone containers) are added to the Docker compose file as \texttt{services} and configured as specified (\eg by setting their network aliases and environment variables as indicated in the in-memory application representation).
The automatically generated \textit{toskosed} images are then used to implement the services corresponding to containers hosting some application component, while the Docker images originally indicated in the TOSCA application specification are used to implement those corresponding to standalone containers.
In addition, Docker volumes are included in accordance to the in-memory application representation, and a Docker overlay network is set for allowing the deployment of the application in both single-host and multi-host infrastructure.
The resulting Docker Compose file follows the schema shown in Fig.~\ref{fig:docker-compose-file}.

\begin{figure}
\scriptsize
\begin{Verbatim}[frame=single]
---
version: '3.7'

networks:
  toskose-network:                      # DOCKER NETWORK DEVOTED TO THE APPLICATION 
    driver: "overlay"                   # setting the network to be an overlay network
    attachable: true

services:                                   

  toskose-manager:                      # CONTAINER HOSTING THE TOSKOSE MANAGER
    image: <toskosed_image>             # setting starting image to corresponding toskosed image 
    networks:                               
      toskose-network:                  # attaching the container to the overlay network
        aliases:                            
        - <alias>                       # setting the alias of the container on the overlay network
    environment:
    - TOSKOSE_MANAGER_PORT=<port>       # setting env. vars. needed by the Toskose Manager
    - TOSKOSE_APP_MODE=<mode>
    - SECRET_KEY=<secret_key>
    ports:                                  
    - <api_port_mapping>                # setting port mapping to expose the RESTful API to users

  <container_node_name>:                # CONTAINER HOSTING SOME APPLICATION COMPONENT 
    image: <toskosed_image>             # setting starting image to corresponding toskosed image 
    init: true                          # enabling Tini as init process
    networks:                               
      toskose-network:                  # attaching the container to the overlay network
        aliases:                        
        - <alias>                       # setting the alias of the container on the overlay network
    volumes:                            
    - <volume_mapping>                  # attaching container to specified volumes (if any)
    environment:                            
    - SUPERVISORD_ALIAS=<alias>         # setting env. vars. for the Unit (i.e., Supervisor instance)
    - SUPERVISORD_PORT=<http_port>
    - SUPERVISORD_USER=<user>
    - SUPERVISORD_PASSWORD=<password>
    - SUPERVISORD_LOG_LEVEL=<log_level>
    - ...                               # setting other env. vars. needed by the hosted components
    ports:                                  
    - ...                               # setting specified port mappings (if any)

  <standalone_container_node_name>      # STANDALONE CONTAINER
    image: <base_image>                 # setting starting image to that indicated in the TOSCA spec 
    networks:                           
      toskose-network:                  # attaching the container to the overlay network
        aliases:                            
        - <alias>                       # setting the alias of the container on the overlay network
    volumes:                            
    - <volume_mapping>                  # attaching container to specified volumes (if any)

volumes:                                # DOCKER VOLUMES (if any)
  ...
\end{Verbatim}
\caption{Schema of the Docker Compose file generated by our prototype implementation of the \toskosepackager.}
\label{fig:docker-compose-file}
\end{figure}

\section{Case study}
\label{sec:case-studies}
We hereby illustrate a case study based on a multi-component application, which designed for testing application orchestrators in practice.
We consider the \thinking application (\url{https://github.com/di-unipi-socc/thinking}), which we developed in the scope of our previous research\cite{fault-aware-management-protocols}, and we show how \toskose enables a component-aware orchestration of such application on Docker Swarm and Kubernetes, in a multi-host setting.

\subsection{The \textit{Thinking} Application}
\label{sec:cs-thinking}
\thinking is an open-source web application allowing its users to share thoughts on a web-based portal, so that other users can read them. 
\thinking is composed by three main components:

\begin{itemize}
    \item A MongoDB database storing the collection of thoughts shared by users. The database is obtained by directly instantiating a MongoDB container, which needs to be attached to a volume where shared thoughts are persistently stored.
    \item A Java-based RESTful Web API to remotely access the database of shared thoughts. The API is hosted on a Maven container, and it requires to be connected to the MongoDB container (for remotely accessing the database of shared thoughts).
    \item A web-based GUI visualising all shared thoughts and allowing to insert new thoughts into the database. The GUI is hosted on a NodeJS container, and it depends on the availability of the API to properly work (as it sends HTTP requests to the API to retrieve/add shared thoughts).
\end{itemize}
For the purposes of this case study, and following the sidecar pattern, the application also includes a Logsniffer in the Maven container running the API of \thinking, which provides a web-based GUI for visualising and filtering the logs of the API.

The GUI, API and Logsniffer are also provided with a set of shell scripts implementing their lifecycle operations.
The operation to install, configure, start, stop and uninstall each of such components are indeed implemented by the scripts \texttt{install.sh}, \texttt{configure.sh}, \texttt{start.sh}, \texttt{stop.sh} and \texttt{uninstall.sh}, respectively.
The API is also equipped with the script \texttt{push\_default.sh}, which can be optionally executed when the API is configured (but not running) to add a default set of thoughts to the MongoDB database. 

\subsection{Modelling \thinking with TOSCA}
\label{ssec:cs-thinking-in-tosca}
\begin{figure}
    \centering
    \includegraphics[scale=\figscale]{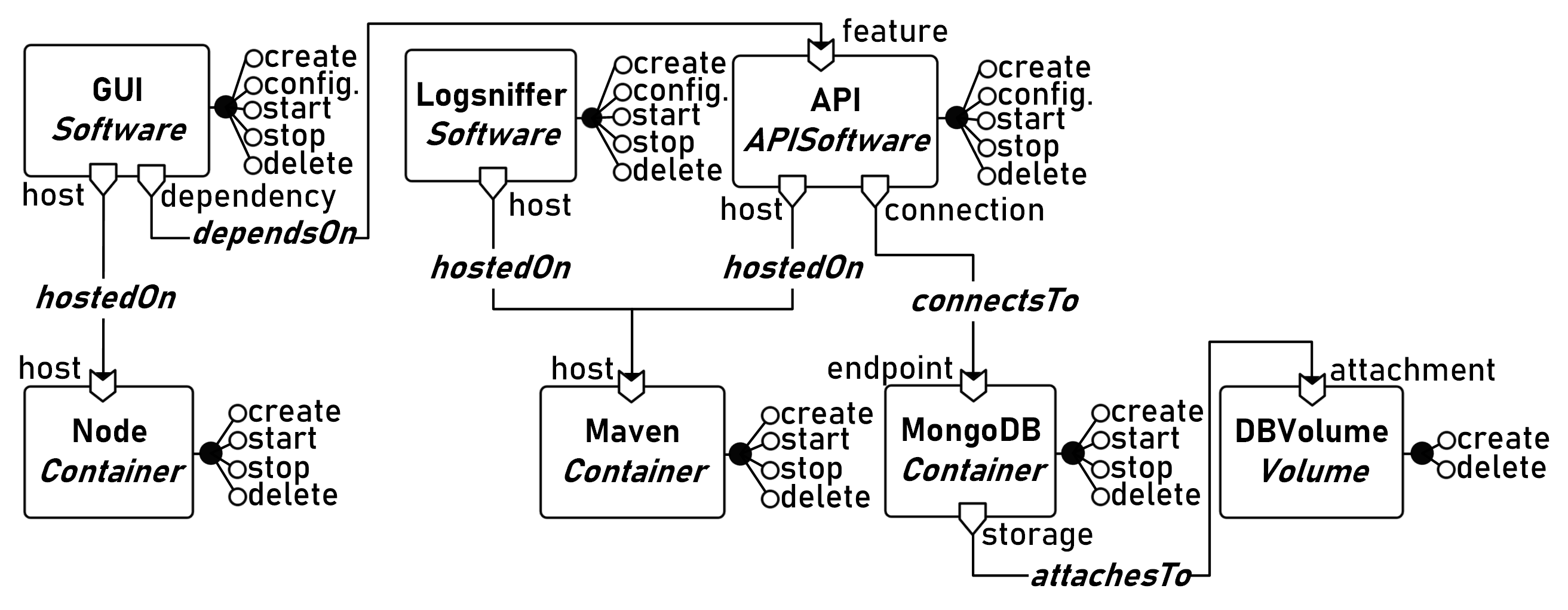}
    \caption{TOSCA-based representation of \thinking, obtained by exploiting the TOSCA types defined by TosKer\cite{tosker}.}
    \label{fig:thinking-in-tosca}
\end{figure}

By exploiting the TOSCA-based representation given by TosKer\cite{tosker} (and recapped in Sect.~\ref{ssec:bg-tosker-types} for making this article self-contained), the \textit{Thinking} application can be modelled in TOSCA as shown in Fig.~\ref{fig:thinking-in-tosca}.
\mongodb is modelled as a component of type \toskerContainer and it is attached to the needed volume (\dbvolume) through a relationship of type \toscaAttachesTo.
\api and \logsniffer are modelled as software components hosted on a component of type \toskerContainer (\ie \maven), with \api being also connected to \mongodb.
Notice that, while \logsniffer is of type \toskerSoftware, \api is of type \apiSoftware, which extends \toskerSoftware to include the \pushdefault operation featured by \api. 
Finally, \gui is modelled as a component of type \toskerSoftware, which is hosted on a component of type \toskerContainer (\ie \node).
The \gui is also indicated as depending on the availability of \api to suitably serve its clients.

The corresponding TOSCA application specification is publicly available on GitHub, in the CSAR packaging such specification together with the artifacts implementing the components of \thinking and their management operations (\url{https://github.com/di-unipi-socc/toskose-packager/blob/master/tests/data/thinking-v2/thinking-v2.csar}).
Concerning artifacts, it is worth noting that the artifact type used to implement \mongodb differ from those of \maven and \node, as \mongodb is qualified to be a standalone container, while \maven and \node{} {are used as system containers for hosting application components}.
In addition, no artifact is provided for implementing the lifecycle operations of the containers and volume in \thinking, as our aim is to piggyback on existing container orchestrators, which natively feature such operations.

\subsection{Generating a Deployable Artifact for \thinking}
\label{ssec:cs-toskosing-thinking}
We generated a Docker Compose file enabling a component-aware orchestration of the management of \thinking on Docker-based container orchestrators by running the \toskosepackager as follows: 
%
\begin{Verbatim}
$ toskose -p thinking.csar toskose.yml
\end{Verbatim}
%
where \texttt{thinking.csar} was a local copy of the CSAR packaging \thinking available on GitHub  (see Sect.~\ref{ssec:cs-thinking-in-tosca}), and \texttt{toskose.yml} was the Toskose configuration file shown in Fig.~\ref{fig:cs-thinking-toskose-config}(a).
\begin{figure}
\centering
\begin{minipage}{.48\textwidth}
\scriptsize
\begin{Verbatim}[frame=single]
nodes:
  maven:
    alias: maven
    port: 9456
    user: user_21ty5
    password: 1t5mYp4ss
    log_level: INFO
    docker:
      name: giulen/thinking-maven-toskosed
      tag: 0.1.3
  node:
    alias: node
    port: 13450
    user: user_a4bc2
    password: p4ssw0rd
    log_level: DEBUG
    docker:
      name: giulen/thinking-node-toskosed
      tag: 2.1.5
manager:
  alias: toskose-manager
  port: 12000
  user: admin_manager
  password: password_manager
  mode: production
  secret_key: my_secret
  docker:
    name: giulen/thinking-manager-toskosed
    tag: latest
\end{Verbatim}
\end{minipage}
\ 
\begin{minipage}{.48\textwidth}
\scriptsize
\begin{Verbatim}[frame=single]
nodes:
  maven:
    alias: maven
    port: 9001
    user: admin
    password: admin
    log_level: INFO
    docker:
      name: giulen/thinking-maven-toskosed
      tag: 0.1.3
      registry_password:
  node:
    alias: node
    port: 9001
    user: admin
    password: admin
    log_level: INFO
    docker:
      name: giulen/thinking-node-toskosed
      tag: 2.1.5
      registry_password:
manager:
  alias: toskose-manager
  port: 10000
  user: admin
  password: admin
  mode: production
  secret_key: secret
  docker:
    name: giulen/thinking-manager-toskosed
    tag: latest
    registry_password:
\end{Verbatim}
\end{minipage}
\\ \smallskip
\begin{minipage}{.48\textwidth}
\centering
(a)
\end{minipage}
\  
\begin{minipage}{.48\textwidth}
\centering
(b)
\end{minipage}
\caption{Toskose configuration files used for generating a deployment of \thinking, with (a)~being manually created and (b)~being automatically generated by the \toskosepackager.}
\label{fig:cs-thinking-toskose-config}
\end{figure}
Fig.~\ref{fig:cs-thinking-toskose-config}(b) instead shows the Toskose configuration file automatically generated by the \toskosepackager, which we obtained by not specifying any Toskose configuration file while running the \toskosepackager
\begin{Verbatim}
$ toskose -p thinking.csar
\end{Verbatim}
In both cases, \texttt{-p} was set to instruct the \toskosepackager to automatically push \textit{toskosed} images to the Docker Hub.

Both runs of the \toskosepackager successfully generated a Docker Compose file for deploying \thinking on a Docker-based container orchestrator.
Fig.~\ref{fig:cs-thinking-compose-file} shows the Docker Compose file obtained by running the \toskosepackager with the Toskose configuration file in Fig.~\ref{fig:cs-thinking-toskose-config}(a).
The file shows that the \toskosemanager is automatically included among the containers to be deployed and how each container hosting some application component is implemented by a \textit{toskosed} image, with the latter being suitably configured to run a \textit{Supervisor} instance as a {\unit} managing the hosted components.

\begin{figure}[!t]
\scriptsize
\begin{Verbatim}[frame=single]
---
version: '3.7'
services:
  maven:
    image: giulen/thinking-maven-toskosed:0.1.3
    init: true
    networks: 
      toskose-network: { aliases: [ maven ] }
    environment:
    - SUPERVISORD_ALIAS=maven
    - SUPERVISORD_PORT=9001
    - SUPERVISORD_USER=user_21ty5
    - SUPERVISORD_PASSWORD=1t5mYp4ss
    - SUPERVISORD_LOG_LEVEL=INFO
    - INPUT_REPO=https://github.com/matteobogo/thoughts-api
    - INPUT_BRANCH=master
    - INPUT_DBURL=mongodb
    - INPUT_DBPORT=27017
    - INPUT_DBNAME=thoughtsSharing
    - INPUT_COLLECTIONNAME=thoughts
    - INPUT_DATA=/toskose/apps/api/artifacts/default_data.csv
    - INPUT_PORT=8080
    ports: [ "8000:8080/tcp" ]
  node:
    image: giulen/thinking-node-toskosed:2.1.5
    init: true
    networks:
      toskose-network: { aliases: [ node ] }
    environment:
    - SUPERVISORD_ALIAS=node
    - SUPERVISORD_PORT=9001
    - SUPERVISORD_USER=user_a4bc2
    - SUPERVISORD_PASSWORD=p4ssw0rd
    - SUPERVISORD_LOG_LEVEL=DEBUG
    - INPUT_REPO=https://github.com/matteobogo/thoughts-gui
    - INPUT_BRANCH=master
    - INPUT_APIURL=localhost
    - INPUT_APIPORT=8000
    - INPUT_APIRESOURCE=thoughts
    ports: [ "8080:3000/tcp" ]
  mongodb:
    image: mongo:3.4
    init: true
    networks:
      toskose-network: { volumes: [ "dbvolume:/data/db" ] }
  toskose-manager:
    image: giulen/thinking-manager-toskosed:latest
    init: true
    deploy: *id001
    networks:
      toskose-network: { aliases: [ toskose-manager ] }
    environment:
    - TOSKOSE_MANAGER_PORT=12000
    - TOSKOSE_APP_MODE=production
    - SECRET_KEY=my_secret
    ports: [ "12000:12000/tcp" ]
networks:
  toskose-network: { driver: "overlay", attachable: true }
volumes:
  dbvolume:
\end{Verbatim}
    \caption{Docker Compose file for deploying \thinking, automatically generated by the \toskosepackager.}
    \label{fig:cs-thinking-compose-file}
\end{figure}

The Docker compose file in Fig.~\ref{fig:cs-thinking-compose-file} is considered hereafter, as the \texttt{docker-compose.yml} file in the rest of our case study.
For the sake of completeness, it is worth highlighting that all activities shown hereafter were successfully executed also with the Docker Compose file obtained by running the \toskosepackager with the Toskose configuration file in Fig.~\ref{fig:cs-thinking-toskose-config}(b).

\subsection{Deploying and Managing \thinking with Docker Swarm}
\label{ssec:cs-thinking-on-swarm}
\begin{figure}
    \centering
    \fbox{\includegraphics[width=\clwidth]{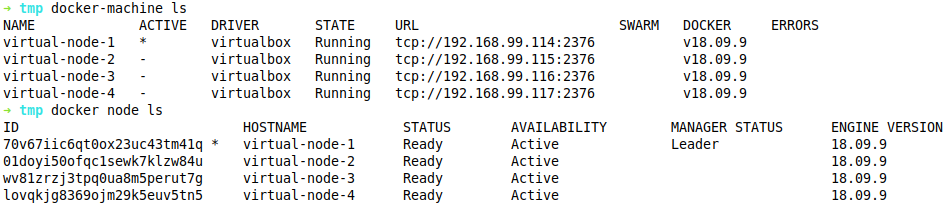}}
    \caption{Multi-host cluster provisioned for allowing the deployment of \thinking with Docker Swarm.}
    \label{fig:cs-thinking-swarm-cluster}
\end{figure}

To deploy the obtained Docker Compose file (\ie \texttt{docker-compose.yml}) with Docker Swarm, we first exploited the Docker Machine tool (\url{https://docs.docker.com/machine}) to create Swarm cluster composed by four virtual machines (Fig.~\ref{fig:cs-thinking-swarm-cluster}).

Following the Docker documentation\cite{docker-stack-deploy}, we then exploited the Docker Stack abstraction for actually enacting the deployment of \thinking on the Swarm cluster.
\begin{Verbatim}
$ docker stack deploy --compose-file docker-compose.yml --orchestrator swarm thinking-stack
\end{Verbatim}
%
\begin{figure}
    \centering
    \fbox{\includegraphics[width=\clwidth]{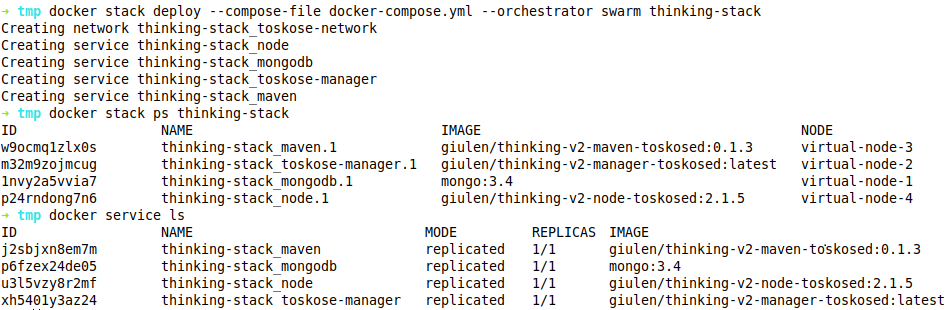}}
    \caption{Execution and outcomes of the command for deploying \thinking on our Swarm cluster.}
    \label{fig:cs-thinking-stack-deploy}
\end{figure}
Fig.~\ref{fig:cs-thinking-stack-deploy} shows the execution and outcomes of the above command.
Docker Swarm cared about spawning the containers in the \thinking application, and of distributing them on the Swarm cluster.
Such containers were however only running their main processes, \ie the \toskosemanager in the case of \texttt{thinking-stack\_toskose-manager}, the \texttt{mongo} application in the case of the standalone container \texttt{thinking-stack\_mongodb}, and the \textit{Supervisor} instances implementing the {\unit}s for the remaining two containers.
The \gui, \api and \logsniffer were instead not deployed yet, as the actual management of their lifecycle was to be orchestrated through the \toskosemanager.

\begin{figure}
    \centering
    \begin{minipage}{.49\textwidth}
        \centering
        \includegraphics[width=\textwidth]{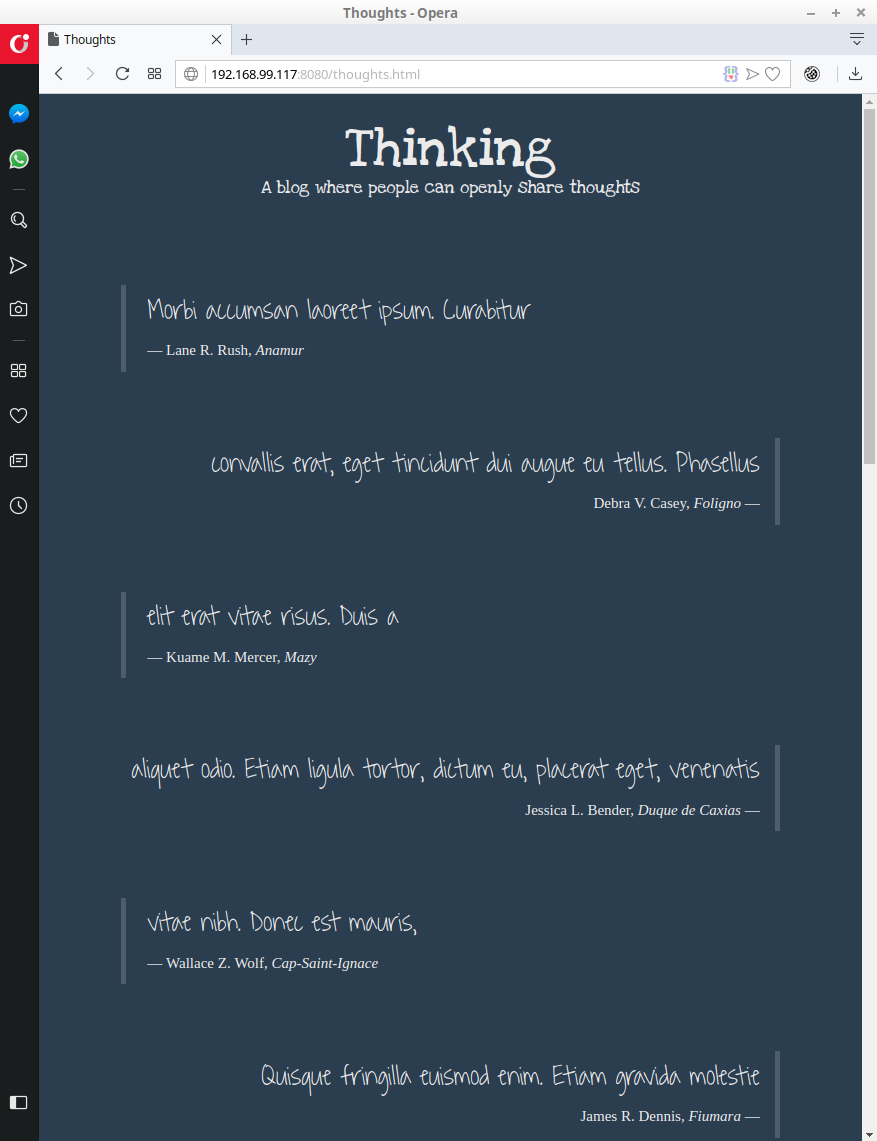}
        (a)
    \end{minipage}
    \ 
    \begin{minipage}{.49\textwidth}
        \centering
        \includegraphics[width=\textwidth]{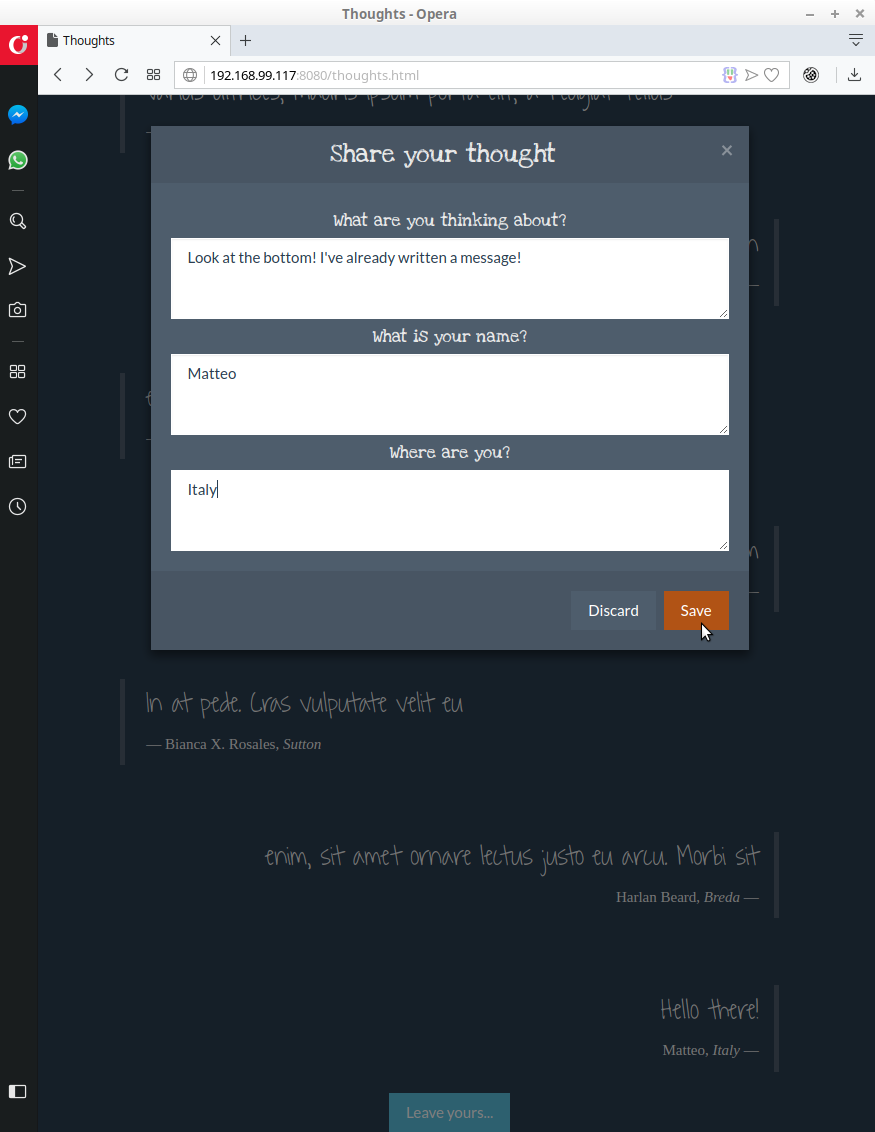}
        (b)
    \end{minipage}
    \caption{Snapshots of the running instance of \thinking obtained after completing its deployment. The snapshots show the web-based interfaces for (a)~reading shared thoughts and for (b)~sharing new thoughts. }
    \label{fig:cs-thinking-running}
\end{figure}

We hence completed the deployment of the \thinking application by exploiting the cURL command-line tool  (\url{https://curl.haxx.se}) for interacting with the \textit{RESTful API} offered by the \toskosemanager.
The latter was running on the virtual machine with IP address \texttt{192.168.99.115} (Figs.~\ref{fig:cs-thinking-swarm-cluster} and \ref{fig:cs-thinking-stack-deploy}), and the API was configured to listen on port \texttt{12000} (Fig.~\ref{fig:cs-thinking-compose-file}).
We hence installed the \api by executing the following command:
\begin{Verbatim}
$ curl -X POST -H "accept: application/json" \  
       "http://192.168.99.115:12000/api/v1/node/maven/api/create"
\end{Verbatim}
Once installed, we configured the \api and instructed it to populate the database with default thoughts by issuing:
\begin{Verbatim}
$ curl -X POST -H "accept: application/json" \
       "http://192.168.99.115:12000/api/v1/node/maven/api/configure"
$ curl -X POST -H "accept: application/json" \
       "http://192.168.99.115:12000/api/v1/node/maven/api/push_default"
\end{Verbatim}
We then started the \api by executing 
\begin{Verbatim}
$ curl -X POST -H "accept: application/json" \
       "http://192.168.99.115:12000/api/v1/node/maven/api/start"
\end{Verbatim}
%
Similarly, we installed and started \logsniffer by executing
\begin{Verbatim}
$ curl -X POST -H "accept: application/json" \
       "http://192.168.99.115:12000/api/v1/node/maven/logsniffer/create"
$ curl -X POST -H "accept: application/json" \
       "http://192.168.99.115:12000/api/v1/node/maven/logsniffer/start"
\end{Verbatim}
and we installed, configured and started the \gui by executing
\begin{Verbatim}
$ curl -X POST -H "accept: application/json" \
       "http://192.168.99.115:12000/api/v1/node/node/gui/create"
$ curl -X POST -H "accept: application/json" \
       "http://192.168.99.115:12000/api/v1/node/node/gui/configure"
$ curl -X POST -H "accept: application/json" \
       "http://192.168.99.115:12000/api/v1/node/node/gui/start"
\end{Verbatim}
As a result, we were able to reach the web-based portal of \thinking both for visualising shared thoughts and for sharing new thoughts (Fig.~\ref{fig:cs-thinking-running}).

It is worth highlighting how we came to complete the deployment of \thinking. 
We first fully relied on the capabilities of Docker Swarm to spawn and manage the Docker containers in \thinking, and the Docker volume needed by \mongodb.
The \toskosemanager then allowed us to manage the rest of the application at component-level, as we were able to remotely invoke the operations managing the lifecycle of the software components in \thinking.

\begin{figure}
    \centering
    \begin{minipage}{.49\textwidth}
        \centering
        \includegraphics[width=\textwidth]{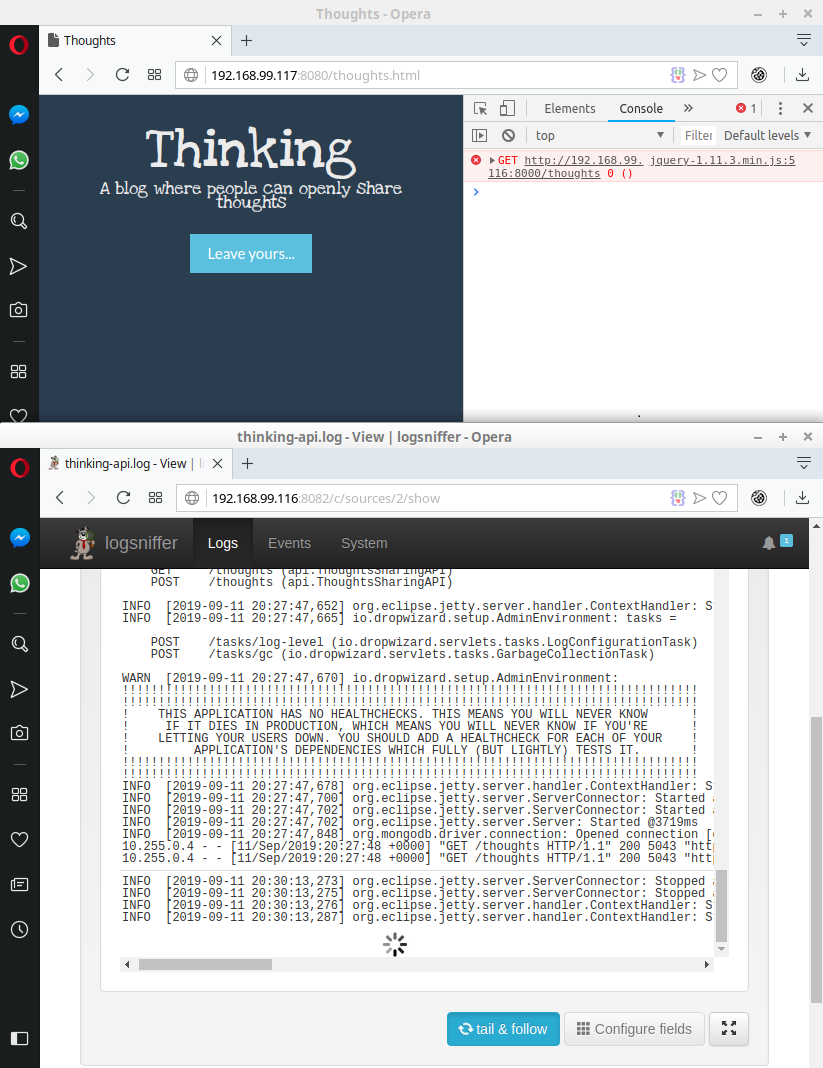}
        (a)
    \end{minipage}
    \ 
    \begin{minipage}{.49\textwidth}
        \centering
        \includegraphics[width=\textwidth]{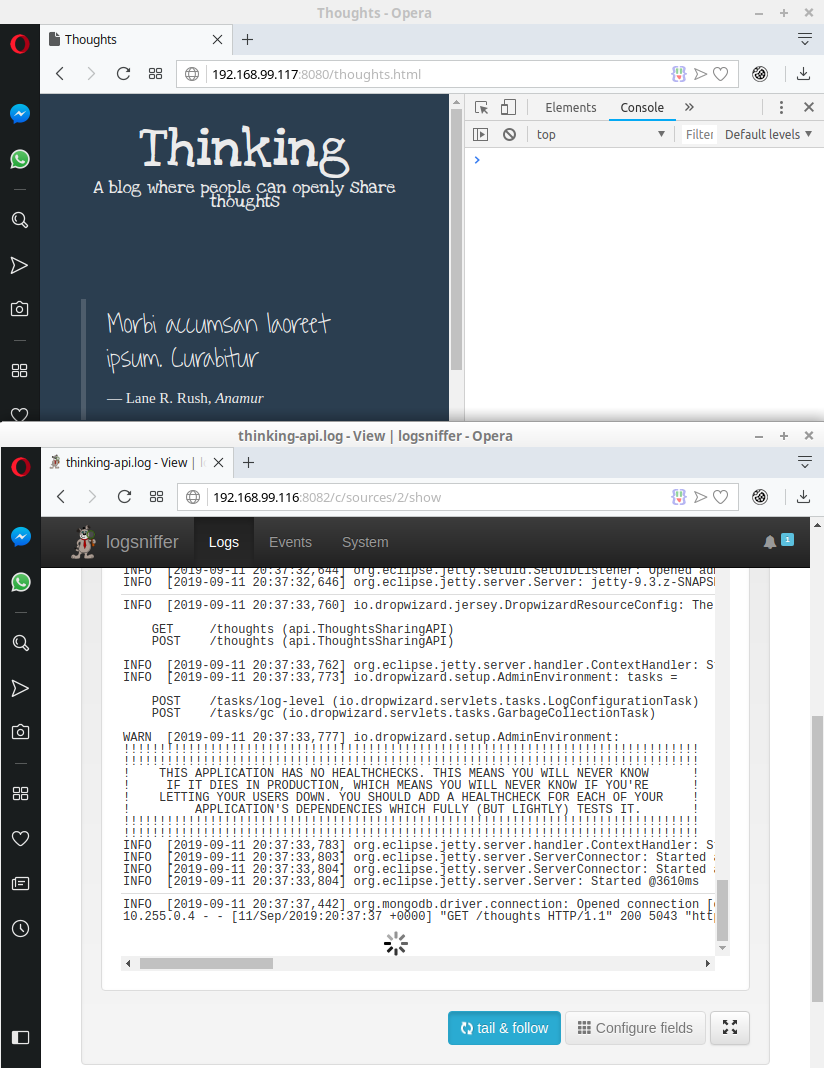}
        (b)
    \end{minipage}
    \caption{Snapshots of the running instance of \thinking and of the logs sniffed by the \logsniffer, (a) after stopping the \gui and (b) after restarting it.}
    \label{fig:cs-thinking-management}
\end{figure}

\smallskip \noindent
To further experiment the component-aware orchestration enabled by our approach with the deployed instance of \thinking, we wished to stop and restart its \api, by also observing the changes actually happening to the application.
We hence stopped the \api of \thinking by remotely invoking its \textit{stop} management operation through the \toskosemanager:
\begin{Verbatim}
$ curl -X POST -H "accept: application/json" \
       "http://192.168.99.115:12000/api/v1/node/maven/api/stop"
\end{Verbatim}
As a result, if connecting to the web-portal shown by \thinking, none of the shared thoughts was displayed.
Fig.~\ref{fig:cs-thinking-management}(a) shows the reason for this, with the console of the browser notifying the failure of the GET request sent to the \api for retrieving shared thoughts.
Such an error is due to the fact that the \api was successfully stopped, as indicated by the logs visualised by the \logsniffer (shown at the bottom of the same figure). 

To proceed with our experiment, we restarted the \api, \ie we remotely invoked its management operation \textit{start} through the RESTful API offered by \toskosemanager.
\begin{Verbatim}
$ curl -X POST -H "accept: application/json" \
       "http://192.168.99.115:12000/api/v1/node/maven/api/start"
\end{Verbatim}
Fig.~\ref{fig:cs-thinking-management}(b) shows the outcomes of such an invocation, \ie the web-portal returned to visualise the shared thoughts, and the logs visualised by \logsniffer showed that the \api was successfully restarted and returned serving HTTP requests.

Even if simple, the above experiment further highlights how our approach enables a component-aware orchestration of the management of a multi-component application.
We were indeed able to stop and restart a component (\ie \api), without requiring to stop the container running it or interfering with the other components running on the same container.
\logsniffer continued to run during the whole experiment, allowing us to visualise the logs of the \api.
This also means that the container hosting \logsniffer and \api continued to run, as expected (as its main process is the \textit{Supervisor} instance implementing the {\unit} managing  \api and \logsniffer).

\subsection{Deploying and Managing \thinking with Kubernetes}
\label{ssec:cs-thinking-on-kubernetes}
We run two different experiments for deploying the Docker Compose file obtained from the \toskosepackager (\ie \texttt{docker-compose.yaml}) on Kubernetes, differing on the tool exploited for doing so, \ie Kompose (\url{https://kompose.io/}) and Compose Object (\url{https://github.com/docker/compose-on-kubernetes}).
For the sake of conciseness, we hereafter only report on that based on Kompose.

\begin{figure}
  \fbox{\includegraphics[width=\clwidth]{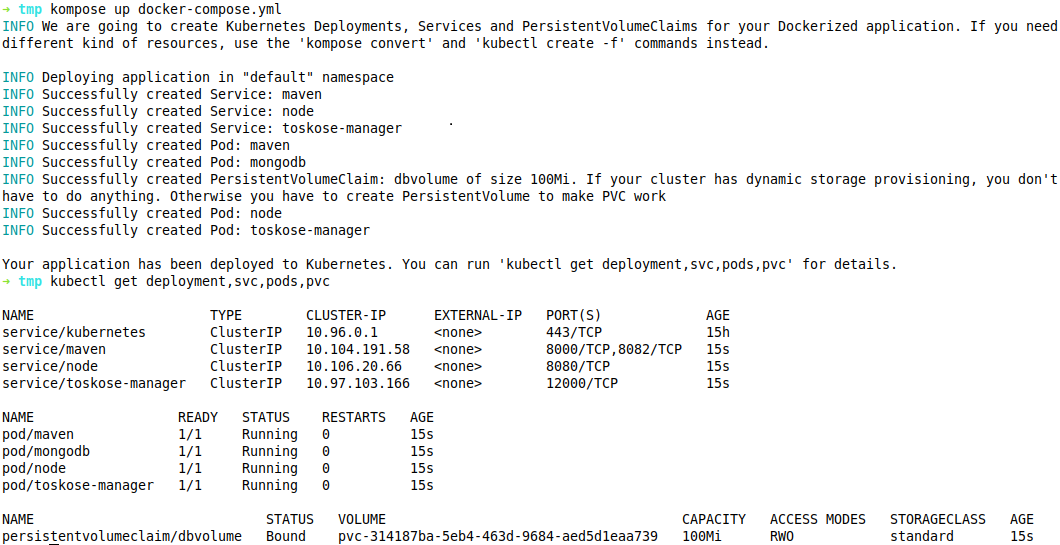}}
  \caption{Execution and outcomes of the command for deploying \thinking on Kubernetes (with Kompose).}
  \label{fig:cs-thinking-on-kompose}
\end{figure}

After creating a Kubernetes cluster, we exploited Kompose to deploy the \texttt{docker-compose.yaml} file on the cluster.
This was done by running
\begin{Verbatim}
$ kompose up docker-compose.yml
\end{Verbatim}
which outcomes are shown in Fig.~\ref{fig:cs-thinking-on-kompose}.
We then exploited the Kubernetes client (\ie \texttt{kubectl}) to specify that the \toskosemanager acts as Ingress node for our deployment\cite{kubernetes-up-running}, hence allowing us to reach the RESTful API it offers.
\begin{Verbatim}
$ kubectl patch svc toskose-manager -p '{"spec":{"type":"LoadBalancer"}}'
\end{Verbatim}
We issued analogous commands to allow remotely accessing \gui and \api, and this completed the deployment and configuration of the containers in \thinking, fully carried out by exploiting existing capabilities featured by the Kubernetes environment.

We then repeated the activities for deploying the software components in \thinking, and for stopping and restarting its \api.
In other words, we repeated the same sequence of \texttt{curl} commands shown in Sect.~\ref{ssec:cs-thinking-on-swarm}, with the only difference given by the IP address where the \toskosemanager was listening (\texttt{10.97.103.166} in this case, as shown in Fig.~\ref{fig:cs-thinking-on-kompose}). 
All such commands successfully executed and resulted in the same outcomes as those presented in Sect.~\ref{ssec:cs-thinking-on-swarm}.

The above, together with the fact that the experiment run by exploiting Compose Object produced the same outcomes, show that we successfully ported the \toskose-based orchestration approach on different Docker-based container orchestrators.
It also shows that, while the management of containers changed accordingly to the employed container orchestrator, the actual management of the components running on such containers remained unchanged, as it was independent from the employed container orchestrator.

\section{Related work}
\label{sec:related}
Various solutions exist for orchestrating the management of multi-component applications, based on TOSCA or Docker.
The closest to ours is TosKer\cite{tosker}, which ---to the best of our knowledge--- is currently the only solution enabling a component-aware orchestration of TOSCA-based application on top of Docker.
It does so by implementing from scratch a new orchestration engine, allowing to coordinate the management of both the software components and the Docker containers forming an application. 
The TosKer engine is designed to run on a single host, which must be configured to provide root privileges to the engine itself (so as to allow it to spawn Docker containers and run application components on them).
Our approach hence differs from that of TosKer, as we enable a component-aware management of TOSCA-based applications on top of existing container orchestrators, which natively support multi-host deployments.
In addition, our approach does not need root priviliges to properly work, hence making it suited also for scenarios where such privileges cannot be granted (\eg on Container-as-a-Service platforms).

Another closely related approach is that tackled by the EDMM modelling and transformation framework\cite{edmm-model,edmm-framework}.
Even if not TOSCA-based, the EDMM modelling and transformation framework allows to specify the software components and Docker containers forming a multi-component application, and the operation allowing to manage each of components and containers\cite{edmm-model}.
It then support the automated generation of the artifacts for deploying the application on top of existing orchestrators, including Docker Compose and Kubernetes\cite{edmm-framework}.
The latter essentially consists in creating deployment scripts coordinating the executable files implementing the management operations of the components of an application, in such a way that the dependencies occurring among such components are satisfied.
It can hence be viewed as a solution for the component-aware deployment of multi-component applications on top of existing deployment platforms.
However, once the deployment is enacted on a container orchestrator, the application is managed through such orchestrator, which considers containers as "black-boxes", \ie not allowing to manage the components forming an application independently from the containers used to host them.
Our approach hence differs from that EDMM-based, as we aim at supporting a component-aware management of multi-component applications during their whole lifecycle.

\smallskip \noindent
Considering containers as "black-boxes" is a baseline also shared by all other existing approaches trying to synergically combine the OASIS standard TOSCA and Docker for orchestrating multi-component applications. 
For instance, Kehrer and Blochinger\cite{tosca-container-mesos} propose to use TOSCA for specifying the internals of a container, which are then manually built by developers to allow their orchestration (as "black-boxes") on top of Mesos.
Our approach is instead intended to enable a component-aware management of multi-component applications on top of existing container orchestrators, by allowing to manage application components independently from their hosting containers.

Other approaches worth mentioning are OpenTOSCA\cite{opentosca}, Alien4Cloud (\url{http://alien4cloud.github.io}), Cloudify (\url{https://cloudify.co}), and the Apache ARIA TOSCA incubator (\url{https://ariatosca.incubator.apache.org}).
OpenTOSCA is an open-source engine for deploying and managing TOSCA applications, which components include containers.
Alien4Cloud, Cloudify and ARIA TOSCA also allow to manage multi-component applications, which components include Docker containers. 
However, they all differ from our approach to managing application since they Docker containers as "black-boxes", \ie not supporting the management of software components separately from that of the containers hosting them.

SeaClouds\cite{seaclouds} and Apache Brooklyn (\url{https://brooklyn.apache.org}) also relate to our approach.
SeaClouds\cite{seaclouds} is a middleware solution for deploying and managing multi-component applications on heterogeneous IaaS/PaaS cloud infrastructure.
SeaClouds fully supports TOSCA, but it lacks a support for Docker containers. The latter makes SeaClouds not suitable to manage multi-component applications including Docker containers.

Apache Brooklyn (\url{https://brooklyn.apache.org}) instead natively supports both TOSCA and Docker containers. 
Thanks to its extension called \textit{Brooklyn-TOSCA}\cite{trans-cloud}, Brooklyn enables the management of the software components and Docker containers forming an application. However, Brooklyn treats Docker containers as black-boxes, and this does not permit managing the components of an application independently of that of the containers used to host them. 

\smallskip \noindent
It is finally worth relating our approach with currently existing solutions for orchestrating multi-container Docker applications.
Docker natively supports their orchestration by means of Docker Compose (\url{https://docs.docker.com/compose}), which allows to indicate the images of the Docker containers forming an application, the virtual network to setup to allow the to intercommunicate, and the volumes to mount to persist their data.
Based on such information, Docker Compose can enact the deployment of the specified application.
Docker Compose however treats containers as "black-boxes", meaning that there is no information on which components are running within a container, and since it does not allow to orchestrate the management of application components independently from their hosting containers.
In addition, no information is provided on the actual interdependencies and interconnections occurring among the components and containers of an application.
Our approach instead allows to explicitly model the software components forming an application, to orchestrate their management independently from their hosting containers, and to explicitly consider the different types of relationships occurring among the components and containers in an application.
This not only makes the interactions occurring among the components of an application easier to understand, but also brings various advantages in terms of reuse and maintenance\cite{toskeriser}.

Other solutions worth mentioning are Docker swarm (\url{https://docs.docker.com/engine/swarm}), Kubernetes (\url{https://kubernetes.io}), and Mesos (\url{http://mesos.apache.org}). Docker swarm permits creating a cluster of replicas of a Docker container, and seamlessly managing it on a cluster of hosts. Kubernetes and Mesos instead permit automating the deployment, scaling, and management of containerized applications over clusters of hosts. Docker swarm, Kubernetes and Mesos differ from our orchestration system as they focus on how to schedule and manage containers (as "black-boxes") on clusters of hosts, while we aim at piggybacking on top of them to enable a component-aware orchestration of the management of multi-component applications.

Similar considerations apply to ContainerCloudSim\cite{containercloudsim}, which provides support for modelling and simulating containerized computing environments. ContainerCloudSim is based on CloudSim\cite{cloudsim}, and it focuses on evaluating resource management techniques, such as container scheduling, placement and consolidation of containers in a data center, by abstracting from the application components actually running in such containers. Our solution instead focuses on allowing to independently manage the components forming an application while delegating container management to a container orchestrator.

\section{Conclusions}
\label{sec:conclusions}
We presented a solution enabling the component-aware management of multi-component application on top of existing Docker-based container orchestrators.
More precisely, starting from an existing TOSCA-based representation for multi-component applications, we illustrated a novel approach allowing to manage the components forming an application independently from the Docker containers used to host them.
We also introduced {three open-source prototype tools} implementing our approach.
These are \toskoseunit (\ie a bundling of \textit{Supervisor} allowing to remotely manage the components running in a container), \toskosemanager (\ie a containerised orchestrator allowing to coordinate the \textit{Supervisor} instances running in the containers of the application), and \toskosepackager (\ie a tool for automatically generating Docker-based deployable artifacts from the TOSCA specification of a multi-component application).

We also illustrated how our approach and {prototype tools} effectively enabled the component-aware management of an existing multi-component application (\ie \thinking) on top of Docker Swarm and Kubernetes.
After representing such applications in TOSCA, we exploited the \toskosepackager for generating deployable Docker Compose files.
The latter were then effectively deployed with both Docker Swarm and Kubernetes on a cluster of virtual hosts.
This allowed us to showcase that the containers forming the infrastructure of the considered applications were deployed and managed by relying on the capabilities of existing Docker-based container orchestrators, while the lifecycle of the software components hosted on such containers was independently orchestrated through the \toskosemanager.
The above held for both considered Docker container orchestrators (\ie Docker Swarm and Kubernetes). 

We believe that this paper can help researchers and practitioners wishing to independently orchestrate the components and containers forming an application.
For instance, we discussed several issues while illustrating the development of our solution, \eg signal management and zombie reaping, or potential conflicts when packaging multiple components in the same container. 
The discussion on issues and their possible solutions can be of help to researchers and practitioners needing to face similar problems while developing alternative solutions to ours, or simply because their applications need multiple components to reside in a single container.

In addition, the \toskose open-source toolset can already be exploited (as-is) by researchers and practitioners to enable a component-aware orchestration of their applications on existing container orchestrators. 
The current prototype of \toskose can also be exploited as the basis for the development
of other research solutions or tools, or for validating existing approaches.
For instance, we exploited \toskose to further validate the outcomes of our former research, \ie we exploited it to run TOSCA application specification automatically completed by \textsc{TosKeriser}\cite{toskeriser}.
The latter is a tool for completing TOSCA application specifications, which automatically discovers and includes the Docker containers offering all what needed to run the components of an application, based on the information automatically retrieved by \textsc{DockerFinder}\cite{docker-analyser}.
Application specification generated by \textsc{TosKeriser} were translated to Docker Compose files, which were then effectively orchestrated on existing Docker-based container orchestrators in a component-aware manner.

\begin{figure}
	\centering
	\includegraphics[scale=\figscale]{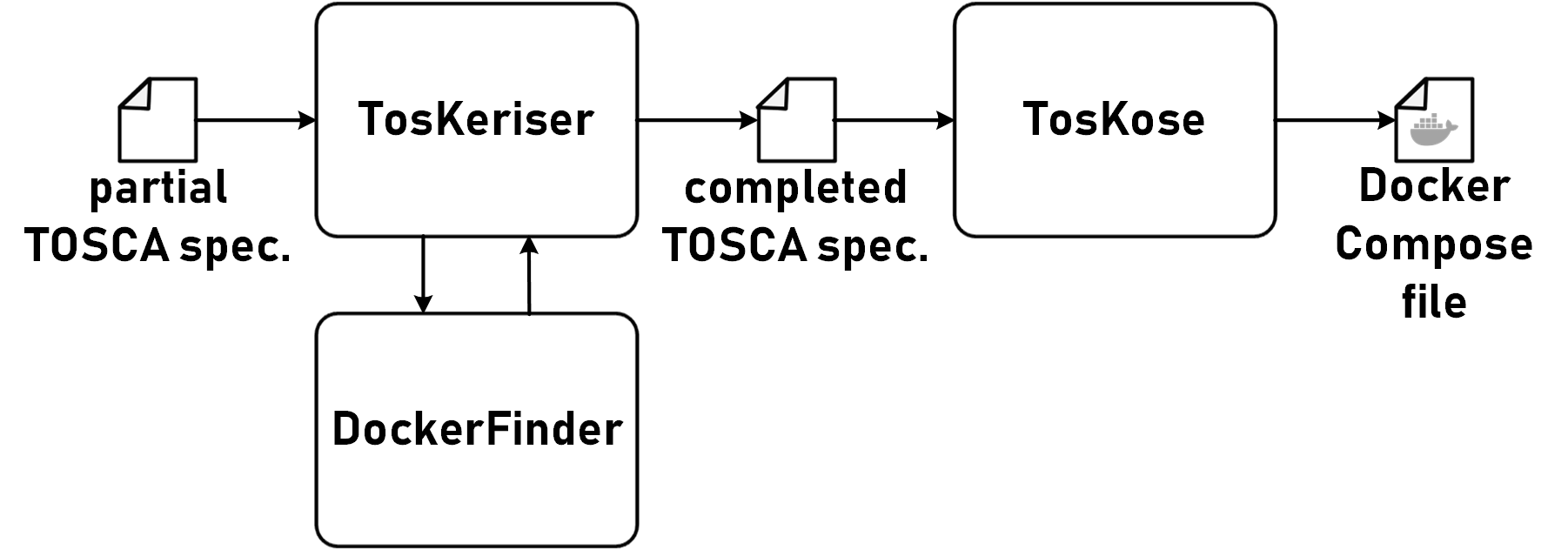}
	\caption{Open-source toolchain for generating deployable solutions from partial TOSCA application specifications, \ie specifications only indicating the components forming an application and the requirements they need to run.}
	\label{fig:toolchain}
\end{figure}

\textsc{TosKeriser}, \textsc{DockerFinder} and \toskose actually form an open-source toolchain, which helps researchers and practitioners in automating the orchestration of multi-component applications with TOSCA and Docker (Fig.~\ref{fig:toolchain}).
They can indeed focus on only describing the components forming an application in TOSCA, and their runtime requirements.
The TOSCA-based application representation is then automatically completed by \textsc{TosKeriser} with the containers allowing to run its components, and then transformed by the \toskosepackager in a deployable solution (\ie a Docker Compose file).
The latter includes the \toskosemanager and the \textit{toskosed} images enabling a component-aware management of the application on top of existing Docker-based container orchestrators.


At the same time, the \toskose open-source {toolset} requires to be further engineered to improve its capabilities.
It currently features basic capabilities for scaling and self-healing components, fully relying on the mechanisms natively featured by the Docker-based container orchestrator employed for deploying an application.
The minimal entity that can be scaled or self-healed is currently a container, and we are currently working on including support for component-aware scaling and self-healing.

We also plan to design and develop a support for automatically determining the workflow of management operations to invoke to allow an application to reach a desired configuration.
Currently, the sequence of operations for reaching a given application configuration is to be manually issued by the application administrator.
We plan to integrate existing analysis and planning techniques (\eg based on Aeolus\cite{aeolus} or on management protocols\cite{fault-aware-management-protocols}), in such a way that the administrator just instructs the \toskosemanager with the desired application configuration, and the \toskosemanager automatically issues the management operations allowing to reach and maintain such a configuration, even in presence of unexpected failures.

\bibliography{src/biblio}%

\smallskip
\noindent 
{All links were last followed on January 27th, 2020.}



\end{document}